\pdfoutput=1
\documentclass[aps,prd,floats,floatfix, twocolumn, amsmath,
superscriptaddress,nofootinbib,showpacs,longbibliography]{revtex4-1}
\usepackage[T1]{fontenc}
\usepackage[utf8]{inputenc}
\DeclareUnicodeCharacter{2009}{\nobreakspace}
\usepackage{lmodern}
\usepackage{times}
\usepackage[varg]{txfonts}
\usepackage[normalem]{ulem}
\usepackage{verbatim}
\usepackage[dvipsnames, usenames]{xcolor}
\definecolor{linkcolor}{rgb}{0.0,0.2,0.5}
\usepackage[hypertexnames=false, unicode, colorlinks=true, linkcolor=linkcolor,
citecolor=linkcolor, filecolor=linkcolor,urlcolor=linkcolor,
pdfusetitle]{hyperref}
\usepackage[all]{hypcap}
\usepackage{graphicx}
\usepackage{xspace}
\usepackage{amssymb}
\usepackage{bm} 
\usepackage{microtype}
\usepackage[english]{babel}
\usepackage{aas_macros}
\usepackage{booktabs}

\allowdisplaybreaks

\usepackage{physics}
\usepackage{orcidlink}
\usepackage{multirow}
\usepackage{enumitem}
\DeclareMathAlphabet{\mathpzc}{OT1}{pzc}{m}{it}

\usepackage{titlesec}

\titleclass{\subsubsubsection}{straight}[\subsubsection]
\newcounter{subsubsubsection}[subsubsection]
\renewcommand\thesubsubsubsection{\thesubsubsection.\arabic{subsubsubsection}}
\titleformat{\subsubsubsection}
  {\normalfont\normalsize\bfseries}{\thesubsubsubsection}{1em}{}
\titlespacing*{\subsubsubsection}{0pt}{3.25ex plus 1ex minus .2ex}{1.5ex plus .2ex}

\begin{document}

\preprint{APS/123-QED}

\title{Line-of-sight Acceleration Driven Effects in Eccentric Compact Binary Coalescences}

\newcommand{\IUCAA}{\affiliation{Inter-University Centre for Astronomy and
    Astrophysics, Post Bag 4, \\ Ganeshkhind, Pune 411007, India}}
\newcommand{\CITA}{\affiliation{Canadian Institute for Theoretical Astrophysics, University of Toronto, 60 St George St,\\ Toronto, ON M5S 3H8, Canada}}  

\author{Mohd Asim Ansari\,\orcidlink{0009-0000-3217-3844}}
\email{asim.ansari@iucaa.in}
 \IUCAA
\author{Avinash Tiwari\,\orcidlink{0000-0001-7197-8899}}
\email{avinash.tiwari@iucaa.in}
\IUCAA
\author{Sajad A. Bhat\,\orcidlink{0000-0002-6783-1840}}
\email{sajad.bhat@iucaa.in}
\IUCAA
\author{Sreejith Nair\,\orcidlink{0000-0002-6428-5143}}
\email{sreejith.nair@iucaa.in}
\IUCAA
\author{Shasvath J. Kapadia\,\orcidlink{0000-0001-5318-1253}}
\email{shasvath.kapadia@iucaa.in}
\IUCAA
\date{\today}

\begin{abstract}
The line-of-sight acceleration (LOSA) effect, imprinted in the gravitational wave (GW) signal of an inspiralling compact binary by an external gravitational potential, is one of the most promising means of probing the environment of compact binaries on a single-event basis. However, this effect is potentially degenerate with other low-frequency effects, particularly the binary's orbital eccentricity. LOSA effects and orbital eccentricity can mimic each other if only one or the other (but not both) are accounted for in the waveform template. In this work, we present the (2, 2) mode, leading eccentric harmonic, frequency-domain phase correction incurred by an accelerating eccentric compact binary, and discuss inconsistencies pertaining to this phase correction in the literature. We show that the degeneracy between LOSA and orbital eccentricity can be resolved if both effects are modeled in the template. To demonstrate this, we inject a GW170817-like signal, with eccentricity and LOSA added to it, and recover the parameters of the injected signal assuming an A+ like noise power spectral density (PSD) of the LVK detector network. We find that the true LOSA and eccentricity parameters, along with other binary parameters, are recovered within the $90\%$ credible interval (CI). This study highlights the importance of incorporating LOSA in eccentric waveform models to robustly and reliably estimate LOSA and eccentricity.
\end{abstract}

\maketitle


\section{Introduction}
The growing catalog of gravitational-wave (GW) detections from compact binary coalescences (CBCs) by the LIGO--Virgo--KAGRA (LVK) network of detectors \cite{TheLIGOScientific:2014jea, TheVirgo:2014hva, KAGRA:2020tym} --- currently numbering nearly 300 up to the second part of the fourth observing run (O4b) \cite{LIGOScientific:2018mvr,LIGOScientific:2020ibl,KAGRA:2021vkt,LIGOScientific:2025hdt,LIGOScientific:2025yae,LIGOScientific:2025slb,LIGOScientific:2026sit,LIGOScientific:2026ifv,LIGOScientific:2026wfs} --- has provided unprecedented access to extragalactic compact binary astrophysics. As detector sensitivities improve \cite{Fritschel2025O5} and longer inspiral signals become accessible, subtle physical effects beyond the standard quasi-circular, isolated-binary paradigm can leave measurable imprints on the observed waveform \cite{Lower:2018seu,Favata:2021vhw,Bhaumik:2024cec}. Properly accounting for such effects is essential for unbiased parameter estimation (PE) and reliable astrophysical interpretation \cite{Favata:2013rwa,Tiwari:2025aec}.

Two effects of particular interest are \emph{orbital eccentricity} and \emph{line-of-sight acceleration} (LOSA). Orbital eccentricity encodes valuable information about the formation channels of compact binaries. While binaries formed through isolated stellar evolution are expected to circularize efficiently before entering the ground-based detector band \cite{Peters1964, Tucker:2021mvo}, dynamically formed systems---such as those in globular clusters, nuclear star clusters, or active galactic nucleus (AGN) disks---may retain measurable eccentricity at observable frequencies \cite{Samsing:2017xmd, Rodriguez:2018pss, Romero-Shaw:2019itr, Samsing:2020tda, Zevin:2021rtf, DallAmico:2023neb}. LOSA, on the other hand, arises when the binary's centre of mass experiences sustained acceleration due to an external gravitational potential, leading to a  Doppler modulation of the GW frequency and an accumulated phase correction over the inspiral \cite{Yunes:2010sm, Bonvin:2016qxr, Inayoshi:2017hgw, Tamanini:2019usx, Gera:2025ugl}.

The impact of LOSA on GW signals from \emph{quasi-circular binaries} has been studied in detail by \cite{Yunes:2011ws, Bonvin:2016qxr, Vijaykumar:2023tjg, Lazarow:2024gdn, Tiwari:2025aec, He:2026bzv, Zhao:2026ydo, Roy:2026mco, Pompili:2026tdf, Roy:2026duh, Gera:2025ugl} (see~\cite{Vijaykumar:2023tjg, 2024arXiv240101743H, Hendriks:2026kys, Roy:2026mco, Pompili:2026tdf, Roy:2026duh, Pathak:2026cik, LIGOScientific:2026qni, LIGOScientific:2026fcf} for LOSA constraints from GW events). \cite{Vijaykumar:2023tjg} derived the phase correction due to LOSA up to 3.5 post-Newtonian (PN) order, and \cite{Tiwari:2025aec} extended the same to include aligned-spin and tidal corrections. \cite{Hendriks:2026kys, Tagawa:2025tfd, Takatsy:2025bfk, Santos:2025ass, Giri:2026wgy, Camilloni:2023xvf, Cocco:2025adu, Cocco:2025udb} and~\cite{Hendriks:2024gpp} have studied this, exploiting dephasing due to the kinematics of the CBC's CoM in the context of three-body systems. \cite{Tiwari:2024pvb} and~\cite{Tiwari:2025qqx} showed\footnote{While this method requires the signal duration to be $\ll$ the outer orbital period,~\cite{Tiwari:2026lvw, Tiwari:2026cnb} have further extended it to a more general case of periodic LOSVs valid for all outer orbital periods.} that the kinematics of the CoM of the CBC imprints itself onto the GW waveform through the imprints of LOSA and other higher-order time derivatives of the line-of-sight velocity (LOSV), and one can profile the merger environments by extracting these imprints from the GW signal. Therefore, their work provides a powerful tool to profile stellar environments on a single-event basis. 

An important subtlety arises when LOSA is inferred using waveform models that assume quasi-circular orbits. The phase modulation induced by LOSA can be partially mimicked by the effects of orbital eccentricity~\cite{Tiwari:2025aec} and vice versa~\cite{Bhat:2025lri}, leading to a degeneracy between these two physical mechanisms. As a result, analyses that neglect eccentricity may misattribute eccentricity to events with LOSA, or vice versa, complicating efforts to extract robust information about the binary's formation history and astrophysical environment. Therefore, systematically exploring the combined impact of orbital eccentricity and LOSA on CBC waveforms becomes important. 

In this work, we address this gap by extending the LOSA framework of~\citep{Vijaykumar:2023tjg} to \emph{eccentric compact binaries}. We derive the GW phase correction due to LOSA for eccentric inspirals and provide the phase correction up to 3 post-Newtonian (PN) higher-order terms relative to the leading-order eccentric correction term within LOSA. Note that ~\cite{Roy:2026mco, Pompili:2026tdf, Pathak:2026cik} have also explored this systematics. ~\cite{Pathak:2026cik} tack-on the quasi-circular LOSA phase correction computed in \cite{Vijaykumar:2023tjg} to the eccentric waveform phase, while  ~\cite{Pompili:2026tdf} self-consistently incorporates LOSA into time-domain waveforms. ~\cite{Roy:2026mco} aims to calculate the same frequency-domain phase correction we model in this work, but does not account for the Doppler shift in the GW frequency, as we show in a subsequent section. The omitted term could lead to significant biases when recovering both LOSA and eccentricity, as we demonstrate with an example injection run.  

Building on our analytic development, we perform Bayesian injection--recovery studies in which both orbital eccentricity and LOSA parameters are included in the injected signal as well as in the inference model. We demonstrate that simultaneous modeling of both these effects significantly alleviates their mutual degeneracy and allows recovery of the true source parameters within the $90\%$  CI. In contrast, neglecting either effect leads to biased PE results.

Our results highlight the importance of jointly modeling intrinsic orbital dynamics and environmental effects in the era of precision GW astronomy. The framework developed here is directly applicable to current LVK observations and will become increasingly relevant for future observing runs and next-generation detectors, where longer inspirals and higher signal-to-noise ratios will further enhance sensitivity to subtle phase corrections.

The remainder of this paper is organized as follows. In Section~\ref{sec:LOSAPhCorrect}, we show the main result of this paper, which is the LOSA-induced phase correction for eccentric binaries, describe its implementation within the underlying eccentric waveform model, and discuss inconsistencies with a related work ~\cite{Roy:2026mco}. Section~\ref{sec:methodology} describes the Bayesian inference framework and injection strategy. In
Section~\ref{sec:results}, we present our results on parameter recovery and degeneracy resolution, and finally in Section~\ref{sec:conclusion}, we provide a concluding discussion.

\section{LOSA-Induced Phase Corrections}\label{sec:LOSAPhCorrect}
Consider a CBC with a total source-frame mass $M_{\rm s}$ and a LOSA of its CoM $a$, such that $\Gamma \equiv a/c$, where $c$ is the speed of light; then its apparent detector-frame mass can be written as
\cite{Bonvin:2016qxr,Vijaykumar:2023tjg}
\begin{multline}
    \label{eq:det_frame_chirp_mass}
    M_{\rm det} = M_{\rm s} (1+z_{\rm cos})(1+z_{\rm Dop})\left(1+\Gamma \,(t_{\rm u} - t_{\rm c})\right) \\ = M \left(1+\Gamma \,(t_{\rm u} - t_{\rm c})\right),
\end{multline}
where $M = M_{\rm s} (1+z_{\rm cos})(1+z_{\rm Dop})$ is the redshifted total mass of the CBC, $z_{\rm cos}$ is the cosmological redshift, $z_{\rm Dop}$ is some constant Doppler shift, $t_{\rm u}$ is the unperturbed time and $t_{\rm c}$ is the time at coalescence in the same frame  --- hereon, we will consider $G = c = 1$, where $G$ is Newton's gravitational constant. As evident from Equation~(\ref{eq:det_frame_chirp_mass}), the detector-frame mass of a binary undergoing LOSA acquires a time dependence, thereby modifying the observed GW signal. 

Under the assumption $\vert (a/c) (t_{\rm u} - t_{\rm c}) \vert \ll 1$, using the stationary phase approximation, we can write the modulated waveform for eccentric CBCs as 
\begin{equation} 
    \label{eq:LOSA_corrected_waveform}
    \tilde{h}_{\rm LOSA}(f) = \tilde{h}(f) \left( 1+\frac{\Delta\mathcal{A}} {\mathcal{A}}(f) \right) \exp \{i \Delta \Psi (f) \},
\end{equation}
where $\tilde{h}(f)$ is the unperturbed waveform --- \texttt{TaylorF2Ecc} waveform model~\cite{Moore:2016qxz} in this case\footnote{We use this as the base waveform for studies throughout this paper.}, $f$ is the GW frequency, $\Delta \mathcal{A} / \mathcal{A}$ is the amplitude correction given by Equation~(4) of~\cite{Ansari:HM_2026}, and $\Delta \Psi$ is the phase correction given by
\begin{equation}
    \label{eq: phase_cor}
    \Delta \Psi (f) =  \frac{25 M \Gamma}{65536 v^{13} \eta ^2} \Biggl[1 - \frac{314 e_0^2}{43} \left( \frac{v_0}{v} \right)^{19/3} + \frac{2983 e_0^2}{731} \left( \frac{v}{v_0} \right)^{5/3} \Biggr],
\end{equation}
at the leading order --- the full correction is given by Equation~\eqref{eq:full_ph_cor} of the Appendix~\ref{app:full_phcor}. Here $v \equiv \left( \pi M f \right)^{1/3}$, $v_0 \equiv \left( \pi M f_0 \right)^{1/3}$, $f_0$ is the reference frequency, $e_0$ is the eccentricity at $f_0$, and the Equations~\eqref{eq: phase_cor} and~\eqref{eq:full_ph_cor} have been derived following~\cite{Vijaykumar:2023tjg}. Specifically, we use the Doppler relations (A1)- (A4) of~\cite{Vijaykumar:2023tjg} to first write
\begin{equation} 
    f_{\rm u} = f_{\rm o} (1+z_{\rm L})\ \ \ \ \ \ \ \ \text{(Doppler shift)},
\label{eq:Doppler_shift}
\end{equation}
\begin{equation}
       dt_{\rm u} = dt_{\rm o} (1+z_{\rm L})^{-1} \ \ \ \text{(Time dilation)}, 
\label{eq:time_dilation}
\end{equation}
\begin{equation}
    v_{\rm u} = v_{\rm o} (1+z_{\rm L})^{1/3},
\label{eq:velocity_transformation}
\end{equation}
and
\begin{equation} 
\label{eq:velocity_evolution_in_observer_frame}
    \frac{dv_{\rm o}}{dt_{\rm o}}= -\frac{\Gamma v_{\rm o}}{3} + (1+\Gamma (t_{\rm u} - t_{\rm c}))^{-\frac{4}{3}}\frac{dv_{\rm u}}{dt_{\rm u}}\,, 
\end{equation}
where $f_{\rm u}$ and $f_{\rm o}$ denote the unperturbed and observed GW frequencies, respectively, $v_{\rm u} = (\pi M f_{\rm u})^{1/3}$, $v_{\rm o} = (\pi M f_{\rm o})^{1/3}$, and $z_{\rm L} = \Gamma (t_{\rm u} - t_{\rm c})$. We then use Equations~(6.10) and (6.7a) of
 ~\cite{Moore:2016qxz} for $dv_{\rm u}/dt_{\rm u}$ (see Equation~(\ref{eq:velocity_evolution_in_source_frame}) of the Appendix \ref{app: derivation}) and $t_{\rm u} - t_{\rm c}$ in Equation~\eqref{eq:velocity_evolution_in_observer_frame}, respectively, by replacing $v \to v_{\rm u}$ and $v_0 \to v_{\rm u,0}$, to obtain $dv_{\rm o} / dt_{\rm o}$ in the observer's frame and invert it to obtain $dt_{\rm o} / dv_{\rm o}$. Subsequently, we integrate it and $\frac{d\phi}{dv_{\rm o}} = \frac{v_{\rm o}^{3}}{M}\, \frac{dt_{\rm o}}{dv_{\rm o}}$ from $v_{\rm o}=v$ to $v_{\rm o}=v_{\rm lso}$, where $v_{\rm lso}=(\pi M f_{\rm lso})^{1/3}$ and $f_{\rm lso}=\frac{1}{6^{3/2}\pi M}$ is the GW frequency at the last stable orbit (LSO), to obtain $t - t_{\rm c}$ and $\phi - \phi_{\rm c}$, respectively. The resulting expressions for $t - t_{\rm c}$ and $\phi - \phi_{\rm c}$ are presented in Equations~(\ref{eq:observer_frame_time}) and (\ref{eq:observer_frame_orbital_phase}), respectively, of the Appendix \ref{app: derivation}. Finally, we use these in Equation
\begin{align}
    \label{eq:fourier-phase}
    \Psi = 2\pi f t(f) - 2\phi(f) -\frac{\pi}{4}\,,
\end{align}
to obtain $\Psi_{\rm LOSA}(f)$ and compute the resulting phase correction using
\begin{align}
    \label{eq:phase_shift}
    \Delta \Psi(f) = \Psi_{\rm LOSA}(f) - \Psi_{\rm No LOSA}(f).
\end{align}

Notice that Equation~\eqref{eq: phase_cor} is different from the phase correction in  ~\cite{Roy:2026mco}, which they obtain to be
\begin{equation}
    \label{eq: phase_cor_soumen}
    \Delta \Psi_{\rm RJ} (f) =  \frac{25 M \Gamma}{65536 v^{13} \eta ^2} \Biggl[1 - \frac{314 e_0^2}{43} \left( \frac{v_0}{v} \right)^{19/3} \Biggr].
\end{equation}
This equation lacks a term present in our formulation. This is because  ~\cite{Roy:2026mco} neglects the Doppler shift in the GW frequency. In section~\ref{soumen_bias_pe}, we highlight the importance of this missing term (the last term in Equation~\eqref{eq: phase_cor}) by showing that it can lead to a systematic bias in the inferred parameters. It can be seen from Equation~\eqref{eq: phase_cor} that the last term becomes more dominant than the second term after $f = (34 / 19)^{3/8} f_0 \approx 1.24 f_0$. This is shown in Figure~\ref{fig:comp_fir_sec_term}. Note that in Appendix ~\ref{sec: rep_our_from_soumen}, we explicitly delineate the procedure to recover the missing term in the phase correction derived in  ~\cite{Roy:2026mco} following the treatment in Section (III) of  ~\cite{Ansari:HM_2026}.

\begin{table}[ht!]
    \centering
    \begin{tabular}{|c|c|}
        \hline 
        \textbf{Parameter} & \textbf{Prior} \\
        \hline \hline
        LOSA $\Gamma \, [{\rm s}^{-1}]$   & $\mathcal{U}$(-0.1, 0.1)   \\
        \hline
        Eccentricity $e_{20}$   & LU ($10^{-5}$, 0.4)   \\
        \hline
        Chirp Mass \, $\mathcal{M} [\rm M_{\odot}]$   & $\mathcal{U}$(0.8, 1.6) \\
        \hline
        Mass ratio $q$   & $\mathcal{U}$(0.1, 1)  \\
        \hline
        Luminosity Distance $D_{\rm L} \, [{\rm Mpc}]$  & PL(10, 2000) $\propto D_{\rm L}^2$ \\
        \hline 
        $\rm RA$ & $\mathcal{U}$(0, $2\pi$) \\
        \hline
        $\rm DEC$ & Cosine($-\frac{\pi}{2}$, $\frac{\pi}{2})$ \\
        \hline
        $\theta_{\rm JN}$ & Sine(0, $\pi$) \\
        \hline
        Polarization angle $\psi$ & $\mathcal{U}$(0, $\pi$) \\
        \hline
        Phase at the reference frequency $\phi$ & $\mathcal{U}$(0, $2\pi$) \\
        \hline
        Geocent time $t_{\rm c} \, [\rm s]$ & 1187008882.0 + $\mathcal{U}$ (0.2, 0.6) \\
        \hline
    \end{tabular}
    \caption{The table of priors used during PE of all injections: $\mathcal{U}$ refers to uniform priors, LU refers to log-uniform priors, while PL refers to power-law priors.}
    \label{tab:priors}
\end{table}

\begin{figure*}[ht!]
    \centering
    \includegraphics[width=0.4\linewidth]{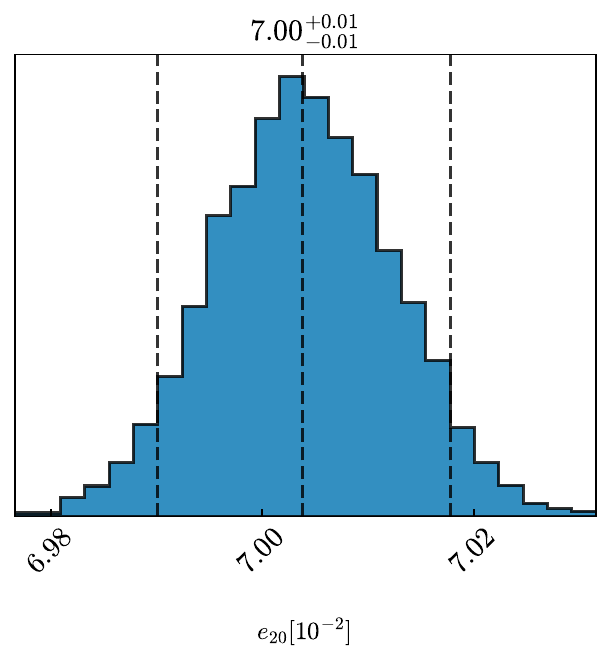}
    \includegraphics[width=0.4\linewidth]{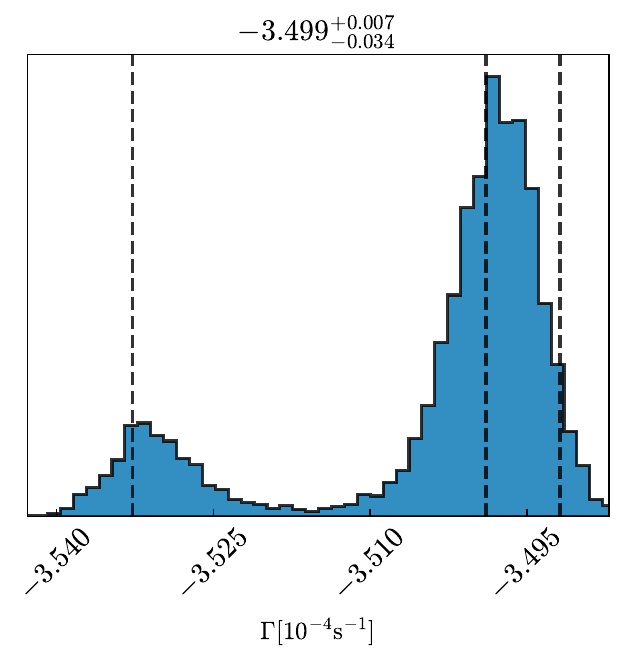}
    \label{fig: losa_ecc_deg}
    \caption{\textbf{LOSA-Eccentricity degeneracy:} The {\it left panel} shows the posterior of eccentricity $e_{20}$ when an accelerated quasi-circular signal with $\Gamma = -5 \times 10^{-5} \, \rm s^{-1}$ is inferred with a vacuum eccentric waveform model. The {\it right panel} shows the posterior of LOSA when a vacuum eccentric signal with $e_{20} = 0.16$ is inferred with an accelerated quasi-circular waveform model. Notice that both posteriors deviate from zero at 90 \% CI.}
\end{figure*}

\section{Injection Setup}\label{sec:methodology}

We employ the LOSA-corrected eccentric waveform model, Equation~\eqref{eq:LOSA_corrected_waveform}, to perform Bayesian PE for a CBC with source properties similar to those of GW170817~\cite{Abbott_2017b} except for the spins and the tidal deformability, which we set to zero across all injections. We choose a GW170817-like system because its relatively low detector-frame chirp mass results in a longer inspiral signal within the detector band, thereby enhancing the accumulated LOSA-induced phase shift and improving the prospects for its measurement.

PE is performed using \texttt{Bilby}~\cite{Ashton_2018jfp} via the \texttt{Bilby\_pipe} pipeline~\cite{Romero-Shaw_2020owr}, and posterior sampling is carried out with the dynamic nested sampler \texttt{Dynesty}~\cite{Speagle_2020}. 
The likelihood is computed using the standard \texttt{GravitationalWaveTransient} likelihood for CBCs over the frequency range $20$--$1792~\mathrm{Hz}$ with a sampling rate of $4096~\mathrm{Hz}$ and a signal duration of 256 s. We set the geocentric time $t_{\rm c}$ to GPS time (1187008882.4) and assume the detector network [H1, L1, V1, K1] operates at O5 sensitivities \footnote{The power spectral densities for this run are taken from \href{https://dcc.ligo.org/LIGO-T2000012-v1/public}{https://dcc.ligo.org/LIGO-T2000012-v1/public}.}. To reduce the dimensionality of the parameter space, we marginalize over the luminosity distance and coalescence phase ~\cite{Thrane_Talbot_2019}. All analyses are performed in zero noise,  while the likelihood is evaluated using the assumed detector noise PSD. In addition, throughout this paper, we set the spins to zero and inject with eccentricities $0.16$ at the reference frequency of $20~\mathrm{Hz}$, unless specified otherwise. We further note that while the \texttt{TaylorF2Ecc} waveform model is found to be accurate for small orbital eccentricities ($e_0 \lesssim 0.1$) for comparable mass systems~\cite{Moore:2016qxz}, it has been tested to return sane values of the waveform for eccentricities as large as $e_0 = 0.5$~\cite{2020ascl.soft12021L}.


\section{Results}\label{sec:results}
Using the setup described in the previous section, we investigate the impact of LOSA on GW PE through the following self-consistency tests and parameter-bias studies.

\begin{figure*}[ht!]
    \centering
        \includegraphics[width=0.475\linewidth]{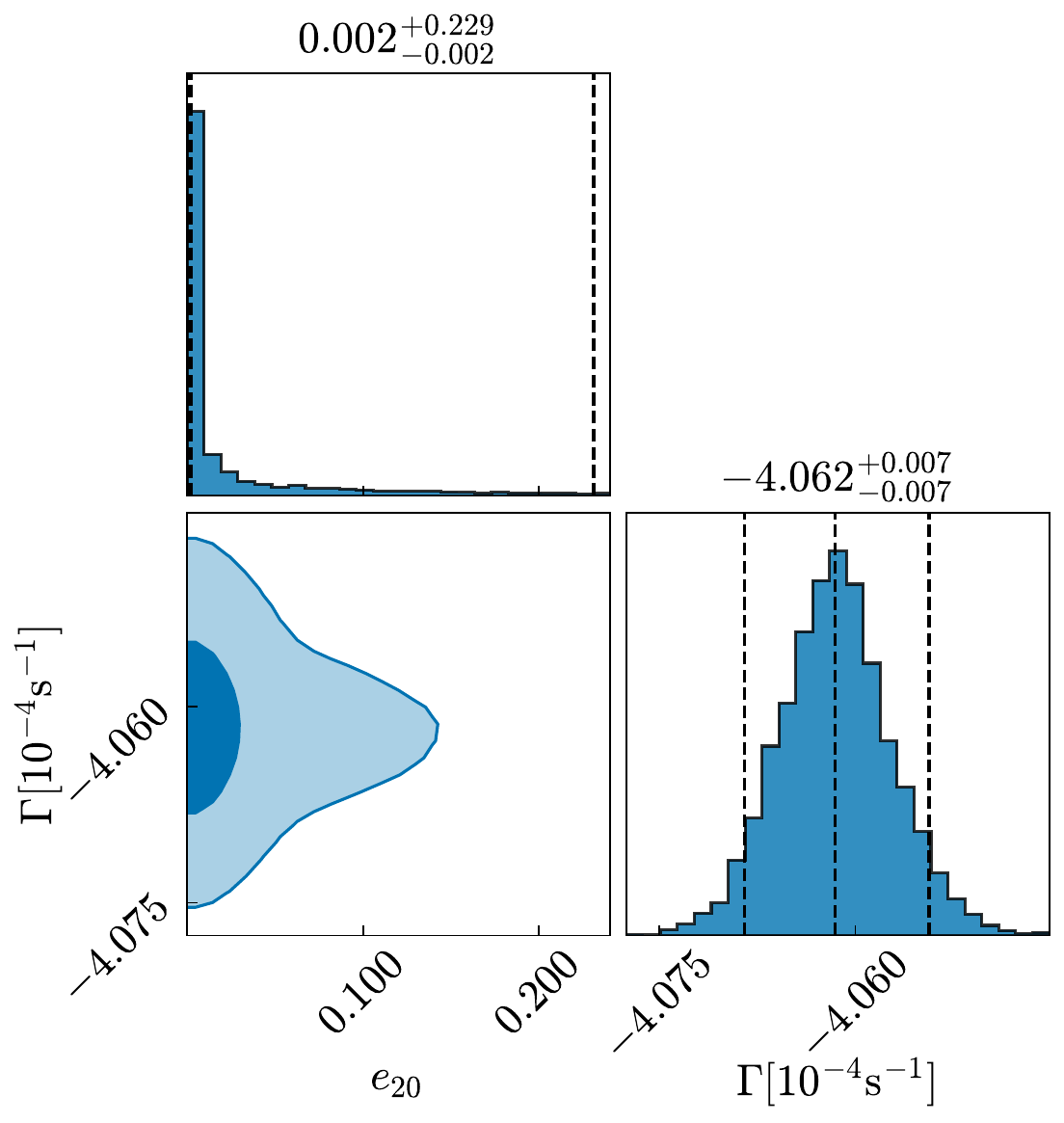}
        \includegraphics[width=0.475\linewidth]{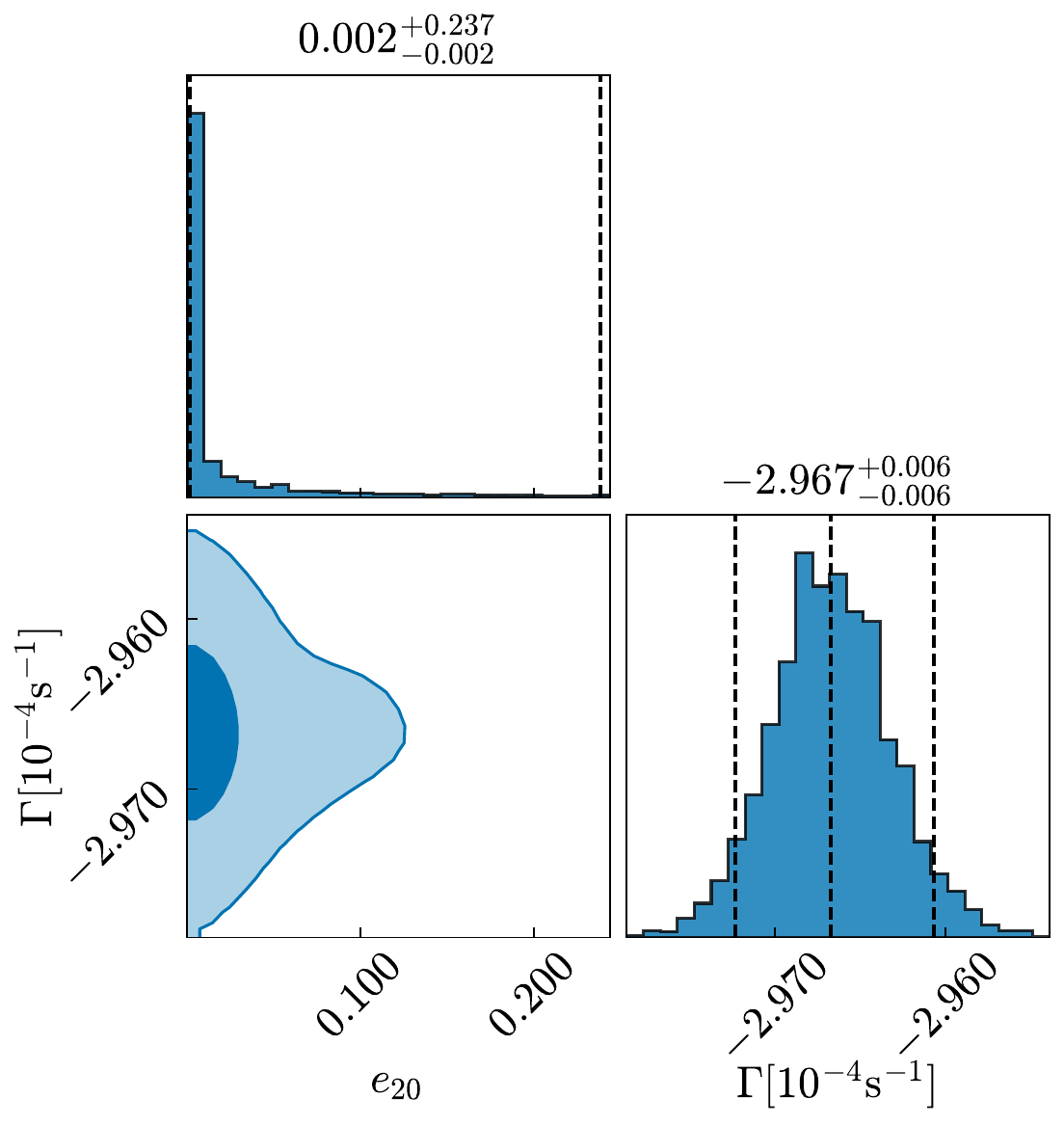}
        \label{fig: losa_ecc_coup}
        \caption{{\bf Bias due to imposing the quasi-circular LOSA corrections:} The {\it left} and {\it right panels} show the joint posteriors of $e_{20}$ and $\Gamma$ for $\{e_{20} = 0.16,\, \Gamma = -5 \times 10^{-5} \, \rm s^{-1} \}$ and $\{e_{20} = 0.16,\, \Gamma = 5 \times 10^{-5} \, \rm s^{-1} \}$, respectively, when the injection model contains eccentric LOSA corrections while the recovery waveform contains only quasi-circular LOSA corrections. Notice that in both cases, the posteriors deviate from their true values.}
\end{figure*}

\subsection{Bias Tests}
Accurate inference of source properties from GW observations requires waveform models that faithfully capture all physically relevant features of the signal. Omitting one or more physical effects from the template waveform can lead to systematic biases in the recovered source parameters as the inference attempts to compensate for the missing physics through other correlated parameters. Since both orbital eccentricity and LOSA introduce characteristic phase modulations in the inspiral waveform, they can partially mimic one another, leading to biased parameter recovery when either effect is neglected. Furthermore, even when both effects are included, an inconsistent treatment of the LOSA correction with respect to the underlying eccentric waveform model can introduce additional systematic biases. In this section, we investigate these two distinct sources of systematic biases through dedicated parameter-estimation studies.

\subsubsection{Eccentricity-LOSA Degeneracy}
\label{sec: mimicry}
This source of systematic bias arises from an incomplete waveform model, where one of the physical effects is omitted during PE. To examine this, we first inject an accelerating ($\Gamma = -5 \times 10^{-5} \, \rm s^{-1}$) quasi-circular CBC in zero noise following Section~\ref{sec:methodology} with \texttt{TaylorF2} as the base model and then perform the PE using \texttt{TaylorF2Ecc} to recover the source parameters. Subsequently, we inject a non-accelerating eccentric ($e_{20} = 0.16$) signal with \texttt{TaylorF2Ecc} and perform PE using a LOSA-modulated waveform with \texttt{TaylorF2} as the base waveform model. Figure~\ref{fig: losa_ecc_deg} shows the corresponding posteriors of eccentricity ({\it left panel}) and LOSA ({\it right panel}). We find that both posteriors exclude zero at the 90\% CI. In addition, from the Bayes factor (BF) perspective, for the eccentric injection with $e_{20} = 0.16$, we find that the accelerating quasi-circular signal is preferred over the non-accelerating one with an $\ln (\rm BF) = 5313.22$; while 
for the accelerating quasi-circular injection with $\Gamma = -5 \times 10^{-5} \rm s^{-1}$, we find that the eccentric signal is favored over a quasi-circular signal by an $\ln  (\rm BF) = 4472.95$. Therefore, mismodeling can lead to biased estimates of the parameters, as also pointed out in ~\cite{Tiwari:2025aec, Bhat:2025lri}.

\subsubsection{Importance of eccentric LOSA corrections}
A second source of systematic bias can arise from an inconsistent treatment of LOSA within the waveform model by excluding the eccentricity-LOSA coupling terms, as done in \cite{Pathak:2026cik}, where quasi-circular LOSA phase corrections derived in \cite{Vijaykumar:2023tjg, Tiwari:2025aec} are simply added to the phase of an eccentric waveform model. To examine this, we first inject a couple of accelerating eccentric CBCs in zero noise with $\{e_{20} = 0.16,\, \Gamma = -5 \times 10^{-5} \, \rm s^{-1} \}$ and $\{e_{20} = 0.16,\, \Gamma = 5 \times 10^{-5} \, \rm s^{-1} \}$ following Section~\ref{sec:methodology} using \texttt{TaylorF2Ecc} as the base model and then perform the PE using only quasi-circular LOSA corrections added on \texttt{TaylorF2Ecc} to recover the source parameters. Figure~\ref{fig: losa_ecc_coup} shows the joint posteriors of $e_{20}$ and $\Gamma$ for the two cases. We find that for both cases, the posteriors of $e_{20}$ and $\Gamma$ deviate from their true values. Therefore, while orbital eccentricity and quasi-circular LOSA corrections are included in such a model, if the LOSA coupling to the eccentric orbital dynamics is omitted, systematic biases in the recovered source parameters can ensue.

\begin{figure}
    \centering
    \includegraphics[width=0.975\linewidth]{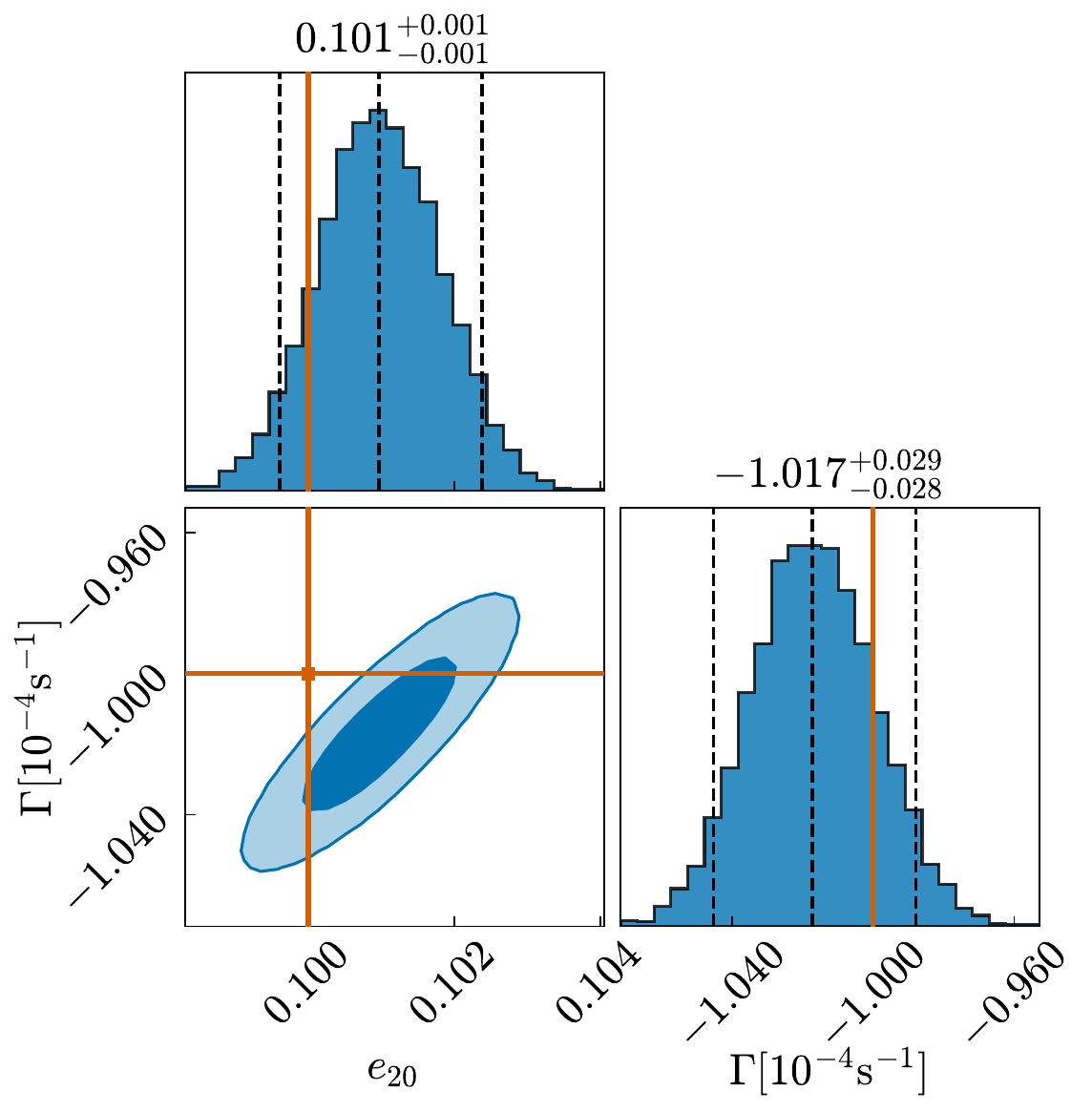}
    \label{fig: bias_pc_RJ}
    \caption{{\bf Bias due to incomplete eccentricity-LOSA coupling terms:} The joint posteriors of $e_{20}$ and $\Gamma$ for $\{e_{20} = 0.1,\, \Gamma = 10^{-4} \, \rm s^{-1} \}$ when the injection model contains eccentric LOSA corrections while the recovery waveform contains incomplete eccentricity-LOSA coupling terms. Notice that while both parameters are recovered within the 90\% CI in $1d$ posteriors, they are not in the $2d$ posterior.}
\end{figure}

\begin{figure}
    \centering
    \includegraphics[width=\linewidth]{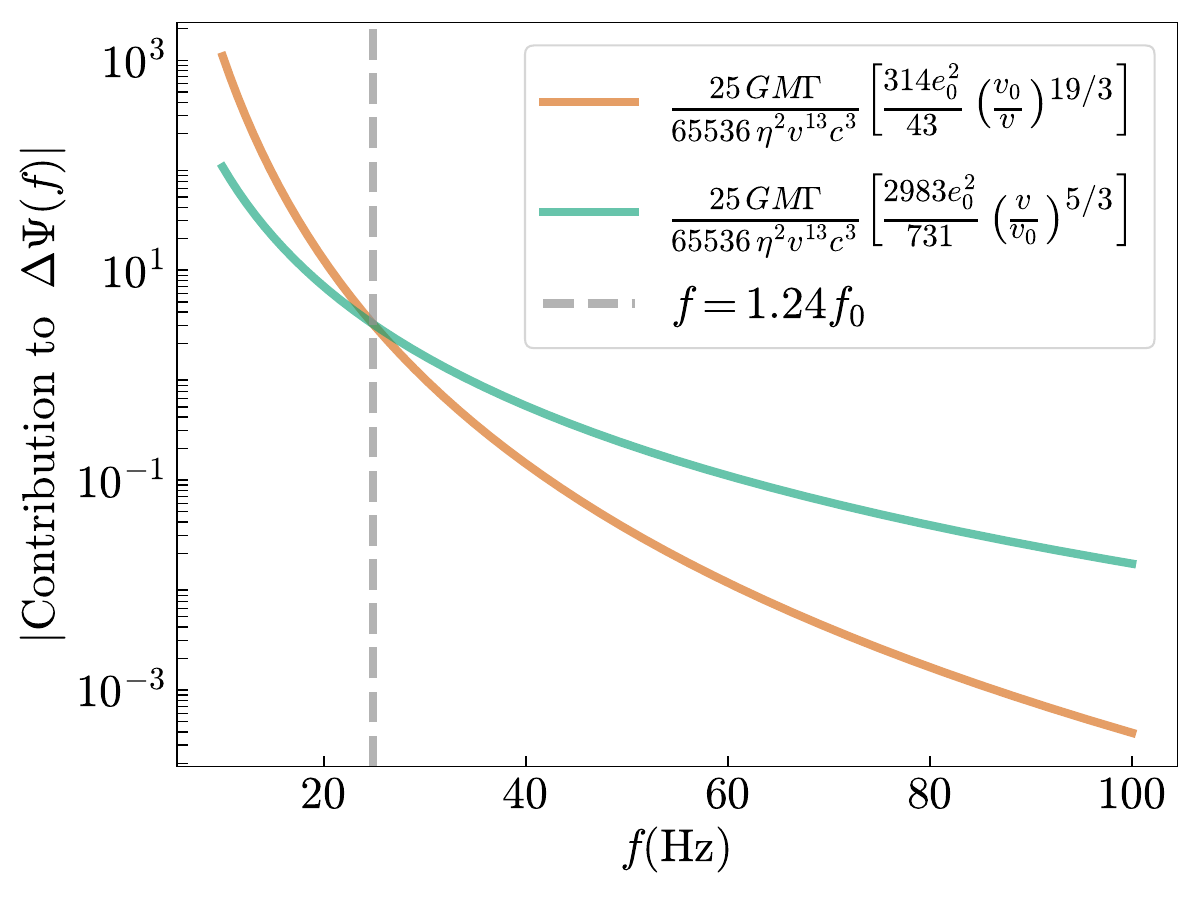}
    \caption{A comparison of the absolute values of the eccentric contributions appearing in Equation~\eqref{eq: phase_cor} for the system considered in Figure~\ref{fig: bias_pc_RJ}. The dashed vertical line shows the limit beyond which the second term takes over the first term for once and for all. This signifies the importance of the missing term in Equation~\eqref{eq: phase_cor_soumen}.}
    \label{fig:comp_fir_sec_term}
\end{figure}

\begin{figure*}
    \centering
    \includegraphics[width=0.475\linewidth]{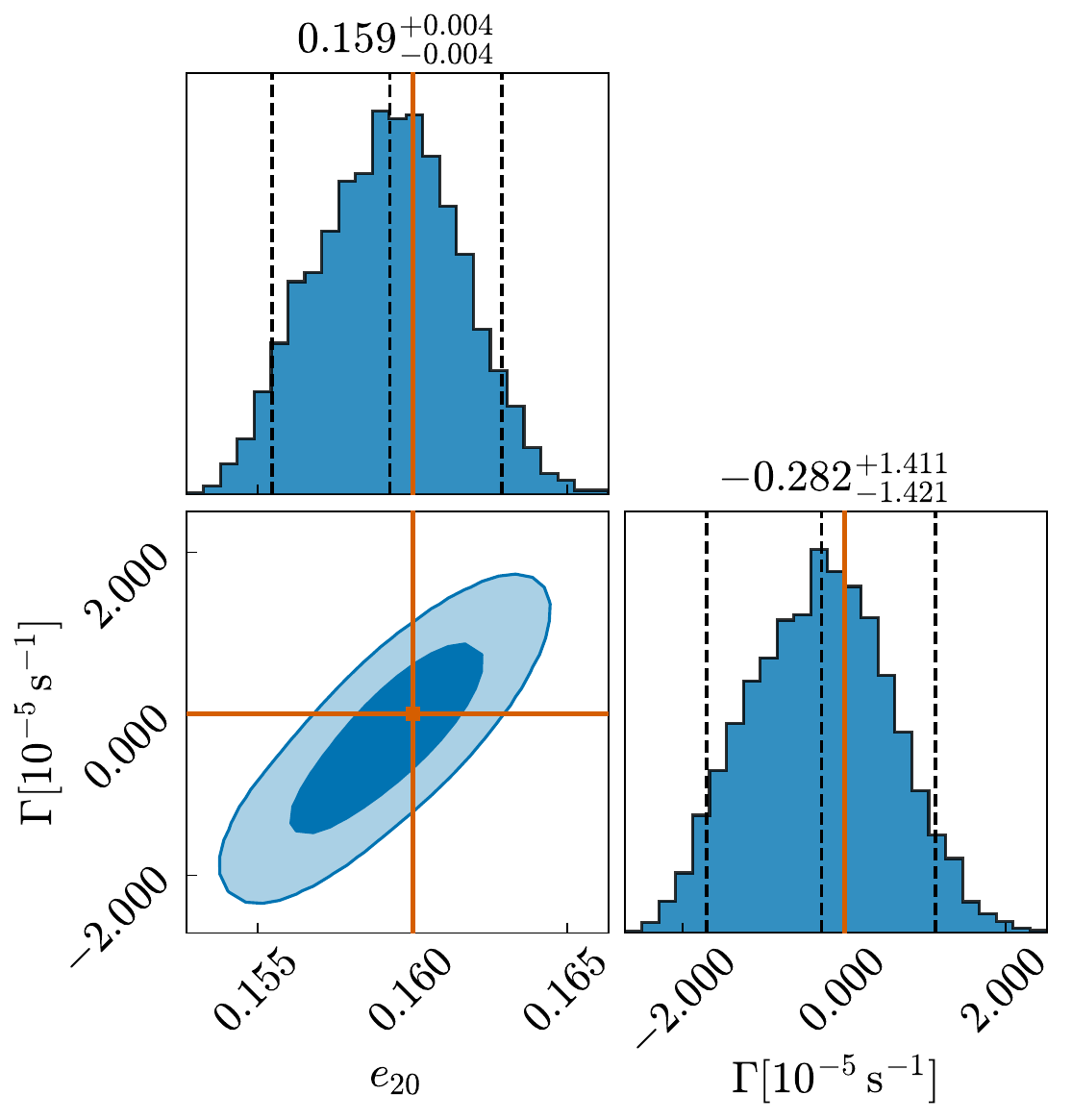}
    \includegraphics[width=0.475\linewidth]{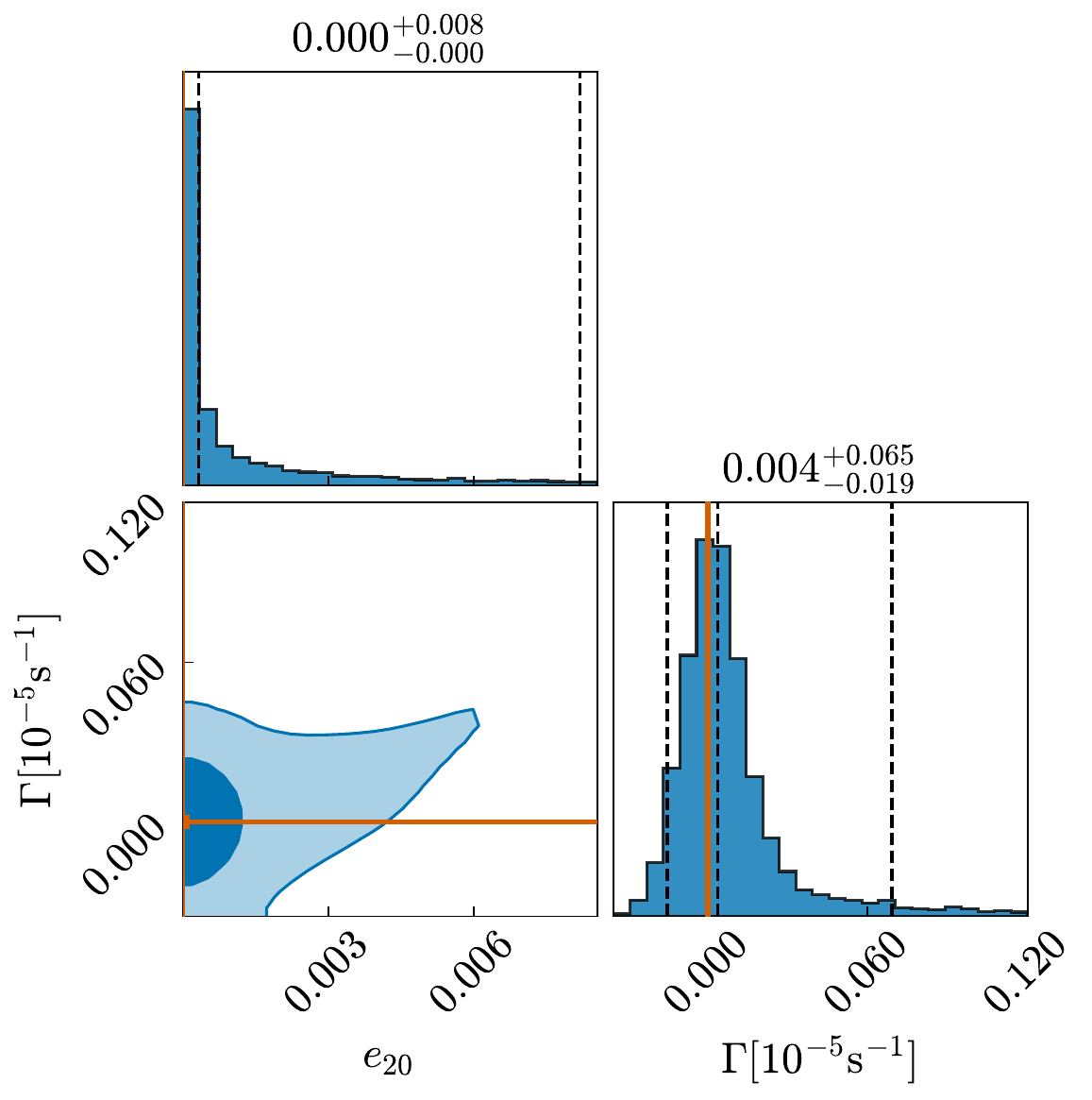}
    \label{fig: self_consist_zero_and_nz}
    \caption{{\bf Removing the Bias:} The {\it left panel} shows the joint posteriors of $e_{20}$ and $\Gamma$ for a non-accelerating eccentric CBC with $\{e_{20} = 0.16 \}$ when the injection and recovery waveforms are the same, while the {\it right panel} shows the same for a non-accelerating quasi-circular  CBC .}
\end{figure*}

\subsubsection{Bias due to incomplete eccentricity-LOSA coupling terms}
\label{soumen_bias_pe}
Another systematic bias can arise from missing some eccentricity-LOSA coupling terms, such as in Equation~\eqref{eq: phase_cor_soumen} where the missing term arises from omitting the Doppler shift to the frequency when calculating the eccentric LOSA phase correction, as done in ~\cite{Roy:2026mco}. To examine this, we first inject an accelerating eccentric CBC in zero noise with  $\{e_{20} = 0.1,\, \Gamma = 10^{-4} \, \rm s^{-1} \}$ following Section~\ref{sec:methodology} using \texttt{TaylorF2Ecc} as the base model and then perform the PE with the same after excluding the third term on the right-hand side of Equation~\eqref{eq: phase_cor}. Figure~\ref{fig: bias_pc_RJ} shows the joint posteriors of $e_{20}$ and $\Gamma$ for this case. We find that while both parameters are recovered within the 90\% CI in the $1d$ marginalized posteriors, they are not recovered within the same interval in the $2d$ posterior. Therefore, including only partial eccentricity-LOSA coupling terms can also lead to biased estimates of the parameters. This highlights the importance of the methodology of computing the LOSA corrections in this paper and in \cite{Vijaykumar} and why the incomplete treatment in~\cite{Roy:2026mco} can lead to non-negligible biases when recovering both parameters. Using \textsc{PyCBC}~\cite{alex_nitz_2024_10473621}, we also calculate the mismatch, minimized over $t_{\rm c}$ and $\phi_{\rm c}$, of the waveform developed in \cite{Roy:2026duh} using their implementation available at \cite{Roy_code} (at the time of writing),   with the waveform developed in this work, employing the O4 PSD. For this, we first set $\Gamma=0$ and notice that for $0 \le e_{20} \lesssim 0.09$ the two non-accelerated eccentric waveforms, namely \texttt{TaylorF2Ecc} and \texttt{pyEFPE} (with the spins set to zero), are in agreement, as the mismatch is smaller than 0.03. But as soon as we turn on LOSA, the mismatch rapidly grows, even for very small values of $\Gamma \sim \mathcal{O}(10^{-6} \rm s^{-1})$ (e.g., mismatch = 0.092 for $e_{20} = 0.05 $ and $\Gamma = - 10^{-6} \rm s^{-1} $). This seems to point towards the absence of the  additional eccentricity LOSA coupling term in the implementation as well.

\begin{figure*}
    \centering
    \includegraphics[width=0.475\linewidth]{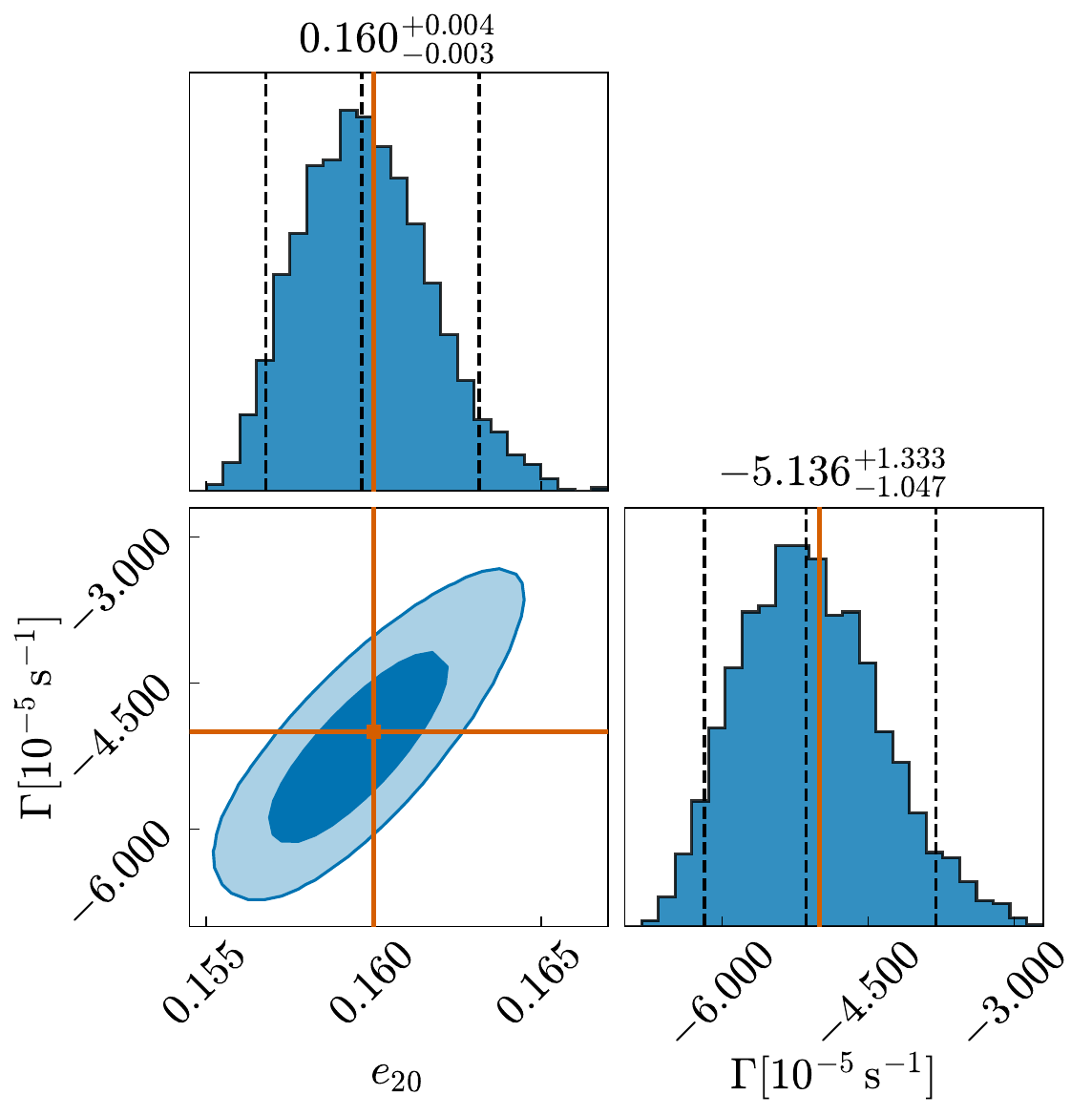}
    \includegraphics[width=0.475\linewidth]{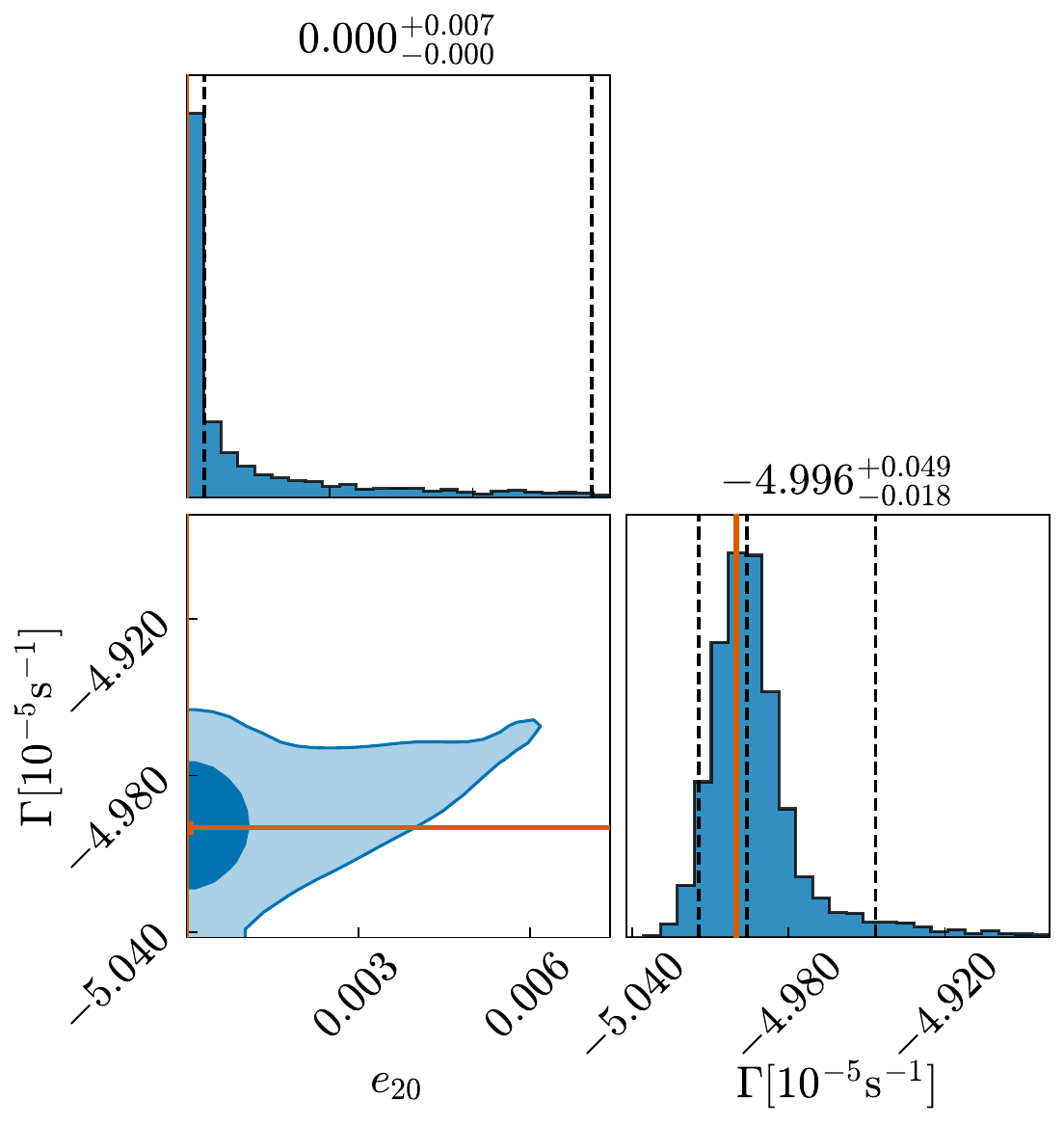}
    \label{fig: self_consist_only_one_nz}
    \caption{{\bf Removing the Bias:} The {\it left panel} shows the joint posteriors of $e_{20}$ and $\Gamma$ for an accelerating eccentric CBC, with $\{e_{20} = 0.16,\, \Gamma = -5 \times 10^{-5} \, \rm s^{-1} \}$, when the injection and recovery waveforms are the same, while the {\it right panel} shows the same for an accelerating quasi-circular CBC with $\{\Gamma = -5 \times 10^{-5} \, \rm s^{-1} \}$.}
\end{figure*}

\subsection{Removing the biases: correct LOSA implementation on an Eccentric Waveform}
\label{sec: self_cons}
We now demonstrate that the LOSA waveform model developed in this paper eliminates the systematic biases discussed in the previous subsections. To this end, we perform PE analyses keeping injection and recovery waveforms the same for four different combinations of the $e_{20}$ and $\Gamma$ values: ({\it i}) when both are zero, and ({\it ii}) when both are non-zero $\{e_{20} = 0.16, \, \Gamma = -5 \times 10^{-5} \, \rm s^{-1} \}$, and ({\it iii}) when only one of them is zero ($\{e_{20} = 0.16, \, \Gamma = 0 \}$ and $\{e_{20} = 0, \, \Gamma = -5 \times 10^{-5} \, \rm s^{-1} \}$). Figure~\ref{fig: self_consist_zero_and_nz} shows the joint posteriors of $e_{20}$ and $\Gamma$ for a non-accelerating eccentric CBC ({\it left panel}) and a non-accelerating quasi-circular CBC ({\it right panel}). Figure~\ref{fig: self_consist_only_one_nz} shows the joint posteriors of the same for an accelerating eccentric CBC ({\it left panel}) and an accelerating quasi-circular CBC ({\it right panel}). For all the injections considered, we find that the recovered values of both parameters are consistent with their injected values and lie within the corresponding $90\%$ CIs, demonstrating that the self-consistent treatment of LOSA and orbital eccentricity eliminates all systematic biases discussed in the previous section. Moreover, we find that the non-accelerating eccentric signal (see the left panel of Figure~\ref{fig: self_consist_zero_and_nz}) is preferred over the accelerating quasi-circular signal (see the right panel of Figure~\ref{fig: losa_ecc_deg})  with an $\ln (\rm BF) = 3293.68$, while the accelerating quasi-circular signal (see right panel of Figure~\ref{fig: self_consist_only_one_nz}) is preferred over the eccentric signal (see the left panel of Figure~\ref{fig: losa_ecc_deg}) with an  $\ln (\rm BF) = 428.17$. Note that for completeness, in Figure~\ref{fig: self_consist_positive_losa} of the Appendix~\ref{app: add_fig}, we also show that both parameters are recovered within their corresponding 90\% CIs when the injected LOSA is positive.

\section{Conclusion}\label{sec:conclusion}
The rapidly increasing sensitivity of current and future GW detectors will require waveform models that include physical effects beyond the standard physics of quasi-circular orbits -- effects that often show up in tandem. Since LOSA and orbital eccentricity may both be present in dynamically formed compact binaries, accurately modelling their combined effects will be essential for unbiased PE, reliable astrophysical inference, and robust tests of general relativity. Degeneracies between different physical mechanisms affecting the GW PE have been noted in earlier studies, including the one between orbital eccentricity and LOSA~\cite{Tiwari:2025aec, Bhat:2025lri}. Such degeneracies complicate the astrophysical interpretation of GW observations, particularly when attempting to infer compact-binary formation channels or environmental properties based on measured deviations from quasi-circular inspirals. In this paper, we have derived the LOSA-induced phase corrections to the frequency-domain phase of GWs from inspiralling eccentric compact binaries within the post-Newtonian framework. The resulting waveform model extends the \texttt{TaylorF2Ecc} approximant by incorporating the effects of LOSA while consistently retaining the eccentric dependence of both the vacuum and acceleration-induced contributions.

Using this waveform model, we investigated the impact of LOSA on GW PE. We showed that LOSA and orbital eccentricity can mimic each other when only one of these effects is included in the recovery waveform, leading to systematic biases in the inferred source parameters. We further demonstrated that an accelerated eccentric signal recovered with a waveform model containing only the quasi-circular LOSA correction \cite{Pathak:2026cik} or incomplete eccentricity-LOSA coupling terms \cite{Roy:2026mco} also yields biased parameter estimates. These highlight the importance of consistently modeling eccentricity and LOSA. We also critically examined the recent LOSA treatment presented in \cite{Roy:2026mco} and identified some non-negligible inconsistencies arising from the omission of the Doppler shift in the GW frequency. These analyses further emphasize the necessity of treating LOSA within a fully self-consistent waveform framework.

To validate the proposed waveform model, we performed PE studies on simulated GW170817-like signals with different combinations of injected LOSA and eccentricity. We find that when both physical effects are incorporated consistently in the recovery waveform, the injected values of the LOSA and eccentricity are recovered accurately and lie within the $90\%$ CI of the posterior distributions. These results demonstrate that the systematic biases identified in this work arise primarily from incomplete waveform modeling rather than limitations of the inference procedure.

Our work has direct implications for formation-channel inference. Orbital eccentricity at detector frequencies is commonly interpreted as evidence for dynamical formation in dense environments such as globular clusters, nuclear star clusters, or AGN disks. However, LOSA naturally arises in these same environments due to the gravitational potential of the surrounding system (see, e.g., \cite{Tiwari:2023cpa}) or the presence of a massive third body (see, e.g., \cite{Tiwari:2026cnb}). Our results show that neglecting LOSA can lead to biased eccentricity estimates, potentially resulting in the misclassification of binaries as intrinsically eccentric. By contrast, joint modeling of LOSA and eccentricity enables the recovery of the true intrinsic orbital parameters, thereby allowing for more reliable formation channel identification.

Beyond formation channels, our findings highlight the importance of jointly interpreting intrinsic binary dynamics and environmental effects. LOSA encodes information about the large-scale gravitational potential in which the binary is embedded, while eccentricity reflects the binary’s dynamical history. The ability to resolve the degeneracy between these effects implies that GW observations can simultaneously constrain both aspects when appropriate waveform models are employed. This strengthens the case for using compact-binary mergers as probes not only of strong-field gravity and stellar evolution, but also of their astrophysical environments (see, e.g., \cite{Tiwari:2024pvb}).

At the population level, unresolved degeneracies between LOSA and eccentricity can systematically bias inferred eccentricity distributions and merger-rate estimates associated with different formation pathways. Our results indicate that incorporating both effects in PE frameworks can mitigate such biases and lead to more robust population-level conclusions, particularly as detector sensitivity improves and smaller phase effects become measurable.

Finally, the resolution of LOSA–eccentricity degeneracies through joint modeling will become increasingly important for next-generation ground-based detectors such as Einstein Telescope (ET) ~\cite{2010CQGra..27s4002P, 2011CQGra..28i4013H} and Cosmic Explorer (CE) ~\cite{2019BAAS...51g..35R, 2022ApJ...931...22S} and future space-based observatories such as DECIGO ~\cite{2017JPhCS.840a2010S, 2011PhRvD..83d4011Y, 2017PhRvD..95j9901Y} , LISA ~\cite{2012CQGra..29l4016A} and TianQin~\cite{2016CQGra..33c5010L}. Longer observation times and higher signal-to-noise ratios will enhance sensitivity to subtle phase modulations, making it essential to account for multiple physical effects simultaneously. Our work demonstrates that doing so not only avoids biased inference, but also unlocks additional astrophysical information encoded in GW signals, underscoring the importance of comprehensive waveform modeling in precision GW astronomy.

\begin{acknowledgements}
  We thank Abhishek Sharma for the LVK’s internal Publications and Presentations review of this work and for his careful reading and valuable feedback on the draft. S.J.K. acknowledges support from SERB grants SRG/2023/000419 and MTR/2023/000086. We are grateful for computational resources provided by the Leonard E. Parker Center for Gravitation, Cosmology and Astrophysics at the University of Wisconsin-Milwaukee, which is supported by National Science Foundation Grants PHY-1626190 and PHY-2110594, the IUCAA LDG cluster Sarathi, and the LIGO Laboratory, a major facility fully funded by the National Science Foundation under Grants PHY-0757058 and PHY-0823459. 
\end{acknowledgements}

\textit{Software}: \texttt{NumPy} \cite{vanderWalt:2011bqk}, \texttt{SciPy} \cite{Virtanen:2019joe}, \texttt{astropy} \cite{2013A&A...558A..33A, 2018AJ....156..123A}, \texttt{Matplotlib} \cite{Hunter:2007}, \texttt{jupyter} \cite{jupyter}, \texttt{LALSuite}~\cite{2020ascl.soft12021L}, \texttt{Bilby}~\cite{bilby_paper}, \texttt{PyCBC}~\cite{alex_nitz_2024_10473621}.

\bibliography{citations}

\onecolumngrid

\appendix

\section*{APPENDIX}\label{appendix}

\section{Intermediate Expressions in the Derivation of the Phase Correction}\label{app: derivation}
Below we present important intermediate expressions used during the derivation of the phase correction.

\noindent
\begin{multline}
    \frac{dv_{\rm u}}{dt_{\rm u}} = \frac{32}{5} \frac{\eta v_{\rm u}^9}{M} \Biggl[
     1 - \Biggl( \frac{743}{336} + \frac{11}{4} \eta \Biggr) v_{\rm u}^2
     + 4\pi v_{\rm u}^3 
     + \Biggl( \frac{34103}{18144} + \frac{13661}{2016} \eta + \frac{59}{18} \eta^2 \Biggr) v_{\rm u}^4 
     - \Biggl( \frac{4159}{672} + \frac{189}{8} \eta \Biggr) \pi v_{\rm u}^5 
     + \Biggl(
        \frac{16447322263}{139708800}
        + \frac{16}{3} \pi^2
         \\ 
    - \frac{1712}{105} \gamma_{\rm E} 
     + \Biggl( \frac{451}{48} \pi^2 - \frac{56198689}{217728} \Biggr) \eta
        + \frac{541}{896} \eta^2
        - \frac{5605}{2592} \eta^3 
     - \frac{856}{105} \ln(16 v_{\rm u}^2)
        \Biggr) v_{\rm u}^6 
     - \Biggl( \frac{4415}{4032}
        - \frac{358675}{6048} \eta
        - \frac{91495}{1512} \eta^2 \Biggr) \pi v_{\rm u}^7
                \\  
    +\frac{157}{24}e_0^{2}\Biggl(\frac{v_{\rm 0u}}{v_{\rm u}} \Biggr)^{19/3} \Biggl \{1 +\Biggl(\frac{2833}{1008} -\frac{197}{36}\eta \Biggr)v_{\rm 0u}^{2} + \frac{377}{72}\pi v_{\rm 0u}{^3} +\Biggl(
    -\frac{1193251}{3048192} - \frac{66317}{9072}\,\eta
    + \frac{18155}{1296}\,\eta^{2}
    \Biggr) v_{\rm 0u}^{4} + \Biggl(\frac{764881}{90720}
                \\ 
    -\frac{949457}{22680}\eta \Biggr)\pi v_{\rm 0u}^{5} + \Biggl(
    \frac{26531900578691}{168991764480}
    - \frac{3317}{126}\gamma_{\rm E}
    + \frac{122833}{10368}\pi^2
    +\Biggl(\frac{9155185261}{548674560} 
    - \frac{3977}{1152} \pi^2 \Biggr)\eta - \frac{5732473}{1306368}\eta^2 - \frac{3090307}{139968}\eta^3 
                    \\ 
    +\frac{87419}{1890}\text{log(2)} 
    - \frac{26001}{560}\text{log(3)} -\frac{3317}{252}\ln(16v^2_{\rm {\rm 0u}})\Biggr)v^6_{\rm {\rm 0u}} 
    + \Biggl ( -\Biggl(\frac{41539}{22608}+ \frac{5413}{5652}\eta \Biggr)+\Biggl(-\frac{117679987}{22788864}+\frac{10486813}{1424304}\eta + \frac{1066361}{203472}\eta^2 \Biggr)v_{\rm 0u}^{2} 
                    \\ 
    - \Biggl(\frac{15660203}{1627776} + \frac{2040701}{406944}\eta \Biggr)\pi v_{\rm 0u}^{3} +\Biggl(
    \frac{49566453289}{68913524736}
    + \frac{237857384155}{17228381184}\eta
    - \frac{1281029377}{68366592}\eta^2
    - \frac{98273015}{7324992}\eta^3
    \Biggr) v_{\rm 0u}^{4}\Biggr)v_{\rm u}^2 + \Biggl(\frac{24871}{11304}\pi 
                    \\ 
    + \Biggl(
    \frac{10065649}{1627776}
    - \frac{4899587}{406944}\,\eta
    \Biggr)\pi v_{\rm 0u}^{2} + \frac{9376367}{813888}\pi^2 v_{\rm 0u}^{3} \Biggr) v_{\rm u}^3 
    + \Biggl(\Biggl(-\frac{122085949}{239283072} 
    + \frac{2133953}{712152}\eta
    - \frac{36497}{101736}\eta^2 \Biggr) + \Biggl(-\frac{345869493517}{241197336576} 
        \\ 
    + \frac{96596798141}{8614190592}\eta - \frac{9444185}{542592}\eta^2 
    + \frac{7189909}{3662496}\eta^3 \Biggr)v_{\rm 0u}^{2} \Biggr)v_{\rm u}^4
    + \Biggl(
    \frac{215395661}{28486080}
    - \frac{8416733}{508680}\,\eta
    \Biggr)\pi v_{\rm u}^{5} +\Biggl(\frac{9765600648106487}{66329267558400}- \frac{2491067}{98910}\gamma_{\rm E}
                \\ 
    -\frac{10610699}{1627776}\pi^2 
    + \Biggl(-\frac{3409129936301}{8614190592} 
    + \frac{2883161}{180864}\pi^2 \Biggr)\eta 
    + \frac{1598264033}{102549888 }\eta^2 -\frac{2773315}{10987488}\eta^3 
    + \frac{5257873}{296730}\text{log(2)} 
                \\ 
    - \frac{1534059}{87920}       \text{log(3)}
    -\frac{2491067}{197820}\ln(16v^2_{\rm u}) \Biggr) v_{\rm u}^6 \Biggr\} \Biggr]
   \tag{A1}
    \label{eq:velocity_evolution_in_source_frame}
\end{multline}


\begin{multline}
t -t_{\rm c} = -\frac{5M}{256\eta v^8} \Biggl[ 1 
    + \Biggl(\frac{743}{252} +\frac{11 \eta}{3} \Biggr)v^2 -\frac{32 \pi v^3}{5} 
    + \Biggl(\frac{3058673}{508032} + \frac{5429 \eta}{504} + \frac{617 \eta^2}{72} \Biggr)v^4 + \Biggl( -\frac{7729 }{252} + \frac{13}{3} \eta \Biggr) \pi v^5 
    + \Biggl( 
    - \frac{10052469856691}{23471078400} 
    \\ 
    + \frac{6848 \gamma_{\rm E}}{105} 
    + \frac{128}{3}\pi^2
    + \frac{3147553127 \eta}{3048192} 
    - \frac{451}{12}\eta \pi^2
    - \frac{15211 \eta^2}{1728} 
    + \frac{25565 \eta^3}{1296} 
    + \frac{6848}{105} \ln(4v)
    \Biggr)v^6 + \Biggl( -\frac{15419335}{127008} - \frac{75703 \eta}{756} 
    \\ 
    + \frac{14809\eta^2}{378} \Biggr) \pi v^7 - \frac{157e_0^{2}v_0^{19/3}}{43v^{19/3}} \Biggl\{1  
    + \Biggl( \frac{2833}{1008}-\frac{197\eta}{36} \Biggr)v_0^{2} + \frac{377}{72}\pi v_0^{3} + \Biggl( - \frac{1193251}{3048192} - \frac{66317 \eta}{9072} + \frac{18155 \eta^{2}}{1296} \Biggr) v_0^{4} + \Biggl(\frac{764881}{90720}
     \\ 
     - \frac{949457 \eta}{22680} \Biggr)\pi v_0^{5} + \Biggl(\frac{26531900578691}{168991764480} + \frac{122833 \pi^{2}}{10368}
    - \frac{3317 \gamma_{\rm E}}{126}
    + \frac{9155185261 \eta}{548674560}
    - \frac{3977 \pi^{2} \eta}{1152} 
    - \frac{5732473 \eta^{2}}{1306368} - \frac{3090307 \eta^{3}}{139968}
    \\ 
    + \frac{87419 \ln(2)}{1890}
    - \frac{26001 \ln(3)}{560}
    - \frac{3317}{252} \ln\!\left(16 v_0^{2}\right)
    \Biggr) v_0^{6}
    + \Biggl(\frac{17592719}{5855472}
    + \frac{1103939 \eta}{209124} + \Biggl( \frac{49840172927}{5902315776} - \frac{42288307 \eta}{26349624}
    \\ 
    - \frac{217475983 \eta^{2}}{7528464}
    \Biggr)v_0^{2} +  \Biggl( \frac{6632455063}{421593984} 
    +\frac{416185003 \eta}{15056928}
    \Biggr) \pi v_0^{3} +\Biggl(-\frac{20992529539469}{17848602906624} 
    - \frac{15317632466765 \eta}{637450103808} + \frac{8852040931 \eta^{2}}{2529563904}
    \\ 
    + \frac{20042012545 \eta^{3}}{271024704}
    \Biggr) v_0^{4} \Biggr)v^2
    + \Biggl(- \frac{2819123 \pi}{384336} + \Biggl(- \frac{7986575459 }{387410688}
    + \frac{555367231 \eta}{13836096}
    \Biggr)\pi v_0^{2} -\frac{1062809371}{27672192}\pi^2 v_0^{3} \Biggr) v^{3} +\Biggl(\frac{955157839}{302766336}
     \\ 
     + \frac{1419591809 \eta}{88306848} + \frac{91918133 \eta^{2}}{6307632}  +\Biggl(\frac{2705962157887}{305188466688}
    + \frac{14910082949515 \eta}{534079816704}
    - \frac{99638367319 \eta^{2}}{2119364352} - \frac{18107872201 \eta^{3}}{227074752}
    \Biggr) v_0^{2} \Biggr)v^4
    \Biggr) 
     \\ 
     + \Biggl(- \frac{166558393 }{12462660}
    - \frac{679533343 \eta}{28486080}\Biggr)\pi v^{5}  +   \Biggl( - \frac{2604595243207055311}{16582316889600000}
    + \frac{924853159 \pi^{2}}{40694400} + \frac{31576663 \gamma_{\rm E}}{2472750}
    + \frac{17598403624381 \eta}{86141905920} 
        \\ 
    - \frac{886789 \pi^{2} \eta}{180864}
    + \frac{203247603823 \eta^{2}}{5127494400}
    + \frac{2977215983 \eta^{3}}{109874880}
    +\frac{226088539 \ln(2)}{7418250} - \frac{65964537 \ln(3)}{2198000} + \frac{31576663 \ln(16 v^2)}{4945500}\Biggr)v^6 \Biggr\}
             \\ 
     + \frac{65M \Gamma}{1536v^8 \eta} 
    \Biggl[ 1 
    + \Biggl( \frac{8173}{1638} 
    + \frac{242 \eta}{39} \Biggr)v^2
    - \frac{128 \pi v^3}{13} + \Biggl( \frac{1755623}{122304} 
    + \frac{32633 \eta}{1092} + \frac{1101 \eta^2}{52} \Biggr) v^4 + \Biggl( -\frac{83228}{1365} 
    - \frac{4592 \eta}{195} \Biggr) \pi v^5
     \\  
     + \Biggl( -\frac{2274117187691}{5029516800} 
    + \frac{13696 \gamma_{\rm E}}{195}   
    + \frac{66304 \pi^2}{975} 
    + \frac{3311653861 \eta}{2830464} 
    - \frac{3157 \pi^2 \eta}{78} + \frac{2030687 \eta^2}{33696} 
    + \frac{35693 \eta^3}{648} 
    + \frac{13696}{195} \ln(4v) \Biggr)v^6
     \\   
       + \Biggl( -\frac{158992529 }{687960} - \frac{1015907  \eta}{4095} + \frac{838 \eta^2}{4095} \Biggr) \pi v^7 - \frac{18212 e_0^{2}v_0^{19/3}}{1677 v^{19/3}} \Biggl\{1+
    \Biggl(\frac{2833}{1008} - \frac{197\eta}{36}\Biggr) v_0^{2} + \frac{377}{72}\pi v_0{^3} + \Biggl( - \frac{1193251}{3048192} - \frac{66317 \eta}{9072} 
    \\  
    + \frac{18155 \eta^{2}}{1296} \Biggr) v_0^{4}
     + \Biggl(\frac{764881}{90720} - \frac{949457 \eta}{22680} \Biggr)\pi v_0^{5} + \Biggl(\frac{26531900578691}{168991764480} + \frac{122833 \pi^{2}}{10368}
    - \frac{3317 \gamma_{\rm E}}{126}
    + \frac{9155185261 \eta}{548674560}
    - \frac{3977 \pi^{2} \eta}{1152}
    \\  
    - \frac{5732473 \eta^{2}}{1306368}
    - \frac{3090307 \eta^{3}}{139968}
    + \frac{87419 \ln(2)}{1890}
    - \frac{26001 \ln(3)}{560} - \frac{3317}{252} \ln\!\left(16 v_0^{2}\right)
    \Biggr) v_0^{6} + \Biggl(\frac{64734553}{12129192}
    + \frac{24319451\,\eta}{3032298}
    \\  
     + \Biggl(\frac{183392988649}{12226225536}
    - \frac{212207749\,\eta}{31839129} - \frac{4790931847\,\eta^{2}}{109162728}
    \Biggr) v_0^{2} + \Biggl(
    \frac{841549189}{30113856}
    + \frac{316152863 \eta}{7528464}
    \Biggr) \pi v_0^{3} +\Biggl(-\frac{77244570101803}{36972106020864}
    \\ 
    - \frac{389631322734485\,\eta}{9243026505216}
    + \frac{1775618540137\,\eta^{2}}{110036029824}
    + \frac{441519632905\,\eta^{3}}{3929858208}\Biggr) v_0^{4}  \Biggr)v^2 +\Biggl( - \frac{1293323983 }{111457440} \pi + \Biggl(
    -\,\frac{523426691977 }{16049871360}
    \\ 
    + \frac{254784824651 \eta}{4012467840}
    \Biggr) \pi v_0^{2} -\frac{16813211779}{276721920} \pi^{2} v_0^{3} \Biggr)v^3 + \Biggl(\frac{227678724398159}{15918545647872}
    + \frac{108515873603\,\eta}{2560898592} + \frac{228119673709\,\eta^{2}}{6768089136}  
    \\ 
     + \Biggl(\frac{645013826219984447}{16045894013054976}
    + \frac{23395745615203055\,\eta}{573067643323392}
    - \frac{935681369766937\,\eta^{2}}{6822233849088} - \frac{44939575720673\,\eta^{3}}{243651208896}
    \Biggr) v_0^{2} \Biggr)v^4 
    \Biggr) 
    \\ 
    +\Biggl(-\,\frac{457836151293337}{7274604193920}
    - \frac{12358466240305 \eta}{207845834112}
    \Biggr) \pi v^5 + \Biggl( -\,\frac{1531975809647354452421}{3939840047848320000} + \frac{77715805463\,\pi^{2}}{1003116960}
    + \frac{385694126\,\gamma_{\rm E}}{7170975}
    \\ 
    + \frac{4334671685651227\,\eta}{4621513252608}
    - \frac{38421305\,\pi^{2}\,\eta}{1311264}+ \frac{1266559392139631\,\eta^{2}}{8527792311360} + \frac{18335858897291\,\eta^{3}}{182738406672}
    + \frac{452177078 \ln(2)}{21512925}
    \\ 
    - \frac{65964537 \ln(3)}{3187100} + \frac{192847063 \ln(16 v^2)}{7170975} \Biggr)v^6 \Biggr\} + \frac{5966 e_0^{2}v^{5/3}}{1677 v_0^{5/3}} 
    \Biggr\{1+
    \left(\frac{42431}{6384} - \frac{2417\eta}{684}\right) v_0^{2} + \frac{2251}{1710}\pi v_0{^3} + \Biggl(\frac{943173641}{57915648} - \frac{26687 \eta}{3024}
    \\
      + \frac{123719 \eta^{2}}{24624} \Biggr) v_0^{4} + \left(-\frac{4740011}{287280} + \frac{817781 \eta}{215460} \right)\pi v_0^{5} + \Biggl(-\frac{51727624818523}{486491443200} + \frac{16107017 \pi^{2}}{984960}
    + \frac{23861 \gamma_{\rm E}}{1710}
    + \frac{10675715266147 \eta}{10424816640}
    \\ 
        - \frac{969773 \pi^{2} \eta}{21888}
    - \frac{109105939 \eta^{2}}{2757888}
    - \frac{3102349 \eta^{3}}{2659392}
    + \frac{170237 \ln(2)}{1890}
    - \frac{962037 \ln(3)}{10640} + \frac{23861}{3420} \ln \left(16 v_0^{2}\right)
    \Biggr) v_0^{6} + \Biggl(\frac{17592719}{5855472}
    \\
    + \frac{1103939\,\eta}{209124}  
    +\left(\frac{746476659889}{37381333248}
    + \frac{12250263163\,\eta}{500642856} - \frac{2668220563\,\eta^{2}}{143040816}
    \right) v_0^{2} + \left(
    \frac{39601210469}{10012857120}
    + \frac{2484966689 \eta}{357602040}
    \right) \pi v_0^{3} 
    \\
    +\left(\frac{16592988834319879}{339123455225856}
    + \frac{720070291976047\,\eta}{12111551972352}
    - \frac{4540513379243\,\eta^{2}}{144185142528}
    + \frac{136578229141\,\eta^{3}}{5149469376}\right) v_0^{4}  \Biggr)v^2 + \Biggl( - \frac{2819123 }{384336} \pi 
    \\
    + \left(
    -\,\frac{119618208013 }{2453601024}
    + \frac{6813820291 \eta}{262885824}
    \right) \pi v_0^{2} -\frac{6345845873}{657214560} \pi^{2} v_0^{3} \Biggr)v^3 + \Biggl(\frac{955157839}{302766336}
    + \frac{1419591809\,\eta}{88306848} + \frac{91918133\,\eta^{2}}{6307632}  
    \\
     + \left(\frac{40528302266609}{1932860289024}
    + \frac{971102392511935\,\eta}{10147516517376}
    + \frac{4838228099263\,\eta^{2}}{120803768064} - \frac{222166127461\,\eta^{3}}{4314420288}
    \right) v_0^{2} \Biggr)v^4 +\Biggl(-\,\frac{166558393}{12462660}
    \\    - \frac{679533343 \eta}{28486080} \Biggr) \pi v^5
    + \Biggl( -\,\frac{2604595243207055311}{16582316889600000} +\frac{924853159\,\pi^{2}}{40694400} 
    + \frac{31576663\,\gamma_{\rm E}}{2472750} + \frac{17598403624381\,\eta}{86141905920} - \frac{886789\,\pi^{2}\,\eta}{180864}  \\  
    + \frac{2977215983\,\eta^{3}}{109874880}
    + \frac{203247603823\,\eta^{2}}{5127494400} 
    +\frac{226088539 \ln2}{7418250} - \frac{65964537 \ln3}{2198000} + \frac{31576663}{4945500} \ln \left(16 v^2 \right) \Biggr)v^6 
    \Biggr\}\Biggr]\Biggr]
    \tag{A2}
    \label{eq:observer_frame_time}
\end{multline}

\begin{multline}
    \phi- \phi_{\rm c} = -\frac{1}{32v^5\eta} \Biggl[1 
    + \Biggl( \frac{3715}{1008} 
    + \frac{55 \eta}{12} \Biggr) v^2 
    - 10 \pi v^3 
    + \Biggl( \frac{15293365}{1016064} + \frac{27145 \eta}{1008} + \frac{3085 \eta^2}{144} \Biggr) v^4  
    - \Biggl(\frac{65
    \eta}{8} - 
    \frac{38645}{672\,\eta}
    \Biggr) \,\ln\!\Biggl(\frac{v}{v_{\mathrm{lso}}}\Biggr) \pi v^5 
      \\  
    + \Biggl (\frac{12348611926451}{18776862720} - \frac{160 \pi^2}{3} - \frac{1712 \gamma_{\rm E}}{21}  
    - \frac{15737765635 \eta}{12192768} + \frac{2255 \pi^2 \eta}{48} 
    + \frac{76055 \eta^2}{6912}  - \frac{127825 \eta^3}{5184} 
    - \frac{1712}{21} \ln(4v) \Biggr)v^6  
      \\  
    +  \Biggl( \frac{77096675}{2032128} + \frac{378515 \eta}{12096} - \frac{74045 \eta^2}{6048} \Biggr) \pi v^7
    -\frac{785 e_0^{2} v_0^{19/3}}{272 v^{19/3}} \Biggl\{1
    + \Biggl(\frac{2833}{1008}- \frac{197\,\eta}{36}\Biggr) v_0^{2} +\frac{377}{72}\pi v_0^{3} + \Biggl(
    - \frac{1193251}{3048192}
    - \frac{66317\,\eta}{9072}
     \\  
    + \frac{18155\,\eta^{2}}{1296}
    \Biggr) v_0^{4} + \Biggl(\frac{764881}{90720} - \frac{949457\eta}{22680} \Biggr)\pi v_0^{5} + \Biggl( \frac{26531900578691}{168991764480}
    + \frac{122833\,\pi^{2}}{10368}
    - \frac{3317\,\gamma_{\rm E}}{126}
    + \frac{9155185261\,\eta}{548674560}
    - \frac{3977\,\pi^{2}\,\eta}{1152}
     \\  
    - \frac{5732473\,\eta^{2}}{1306368}
    - \frac{3090307\,\eta^{3}}{139968}
    + \frac{87419\,\ln(2)}{1890}
    - \frac{26001\,\ln(3)}{560}
    - \frac{3317}{252}\,\ln\!\Biggl(16 v_0^{2}\Biggr)
    \Biggr) v_0^{6} + \Biggl(\frac{6955261}{2215584}+ \frac{436441\,\eta}{79128} 
       \\ 
    + \Biggl(\frac{19704254413}{2233308672}
    - \frac{16718633\,\eta}{9970128}
    - \frac{85978877\,\eta^{2}}{2848608}\Biggr)  v_0^{2} + \Biggl(\frac{2622133397}{159522048} + \frac{164538257 \eta}{5697216} \Biggr) \pi v_0^{3} + \Biggl( -\,\frac{8299372143511}{6753525424128}
     \\ 
    - \frac{6055808184535\,\eta}{241197336576}
    + \frac{3499644089\,\eta^{2}}{957132288}
    + \frac{7923586355\,\eta^{3}}{102549888} \Biggr) v_0^{4} \Biggr)v^2 +
    \Biggl(- \frac{1114537}{141300}\pi 
    + \Biggl(-\frac{3157483321}{142430400} + \frac{219563789 \eta}{5086800} \Biggr)\pi v_0^{2} 
       \\   
    -\frac{420180449}{10173600} \pi^{2} v_0^{3}\Biggr)v^3  + \Biggl(\frac{377620541}{107433216}
    + \frac{561233971\,\eta}{31334688}
    + \frac{36339727\,\eta^{2}}{2238192} +      \Biggl(\frac{1069798992653}{108292681728}
    + \frac{5894683956785\,\eta}{189512193024} 
     \\ 
    - \frac{39391912661\,\eta^{2}}{752032512}
    - \frac{7158926219\,\eta^{3}}{80574912} \Biggr)v_0^{2} \Biggr)v^4 + \Biggl(
    - \frac{131697334}{8456805}
    - \frac{268652717 \eta}{9664920}
    \Biggr)\pi v^{5} + \Biggl(-\frac{204814565759250649}{1061268280934400} 
       \\    
    + \frac{365639621\,\pi^{2}}{13022208}
    + \frac{12483797\,\gamma_{\rm E}}{791280}
    + \frac{34787542048195\,\eta}{137827049472}
    - \frac{8764775\,\pi^{2}\,\eta}{1446912}
    + \frac{80353703837\,\eta^{2}}{1640798208}
    + \frac{5885194385\,\eta^{3}}{175799808}
     \\  
    +\frac{89383841 \ln(2)}{2373840} - \frac{26079003 \ln(3)}{703360} + \frac{12483797 \ln(16 v^2)}{1582560} \Biggr)v^6 
    \Biggr\} 
    + \frac{25M \Gamma}{768v^8 \eta} 
    \Biggl[ 1 
    + \Biggl( \frac{743}{144} 
    + \frac{77 \eta}{12} \Biggr)v^2
    - \frac{52 \pi v^3}{5} + \Biggl( \frac{1755623}{112896} 
       \\ 
    + \frac{32633 \eta}{1008} + \frac{367 \eta^2}{16} \Biggr) v^4 + \Biggl( -\frac{228877}{3360} 
    - \frac{3157 \eta}{120} \Biggr) \pi v^5
     + \Biggl(-\frac{5873342252515}{11266117632} 
    + \frac{1712 \gamma_{\rm E}}{21}   
    + \frac{1184 \pi^2}{15} 
    + \frac{16558269305 \eta}{12192768} 
     \\  
    - \frac{2255 \pi^2 \eta}{48} + \frac{10153435 \eta^2}{145152} 
    + \frac{331435 \eta^3}{5184} 
    + \frac{1712}{21} \ln(4v) \Biggr)v^6 + \Biggl( -\frac{158992529 }{564480} - \frac{1015907  \eta}{3360} + \frac{419 \eta^2}{1680} \Biggr) \pi v^7 
    \\  
    - \frac{10519 e_0^{2}v_0^{19/3}}{1032 v^{19/3}} \Biggl \{1
    + \Biggl(\frac{2833}{1008}- \frac{197\,\eta}{36}\Biggr) v_0^{2} +\frac{377}{72}\pi v_0^{3} + \Biggl(
    - \frac{1193251}{3048192}
    - \frac{66317\,\eta}{9072}
    + \frac{18155\,\eta^{2}}{1296}
    \Biggr) v_0^{4} + \Biggl(\frac{764881}{90720} 
     \\ 
    - \frac{949457\eta}{22680} \Biggr)\pi v_0^{5} + \Biggl(\frac{26531900578691}{168991764480}
    + \frac{122833\,\pi^{2}}{10368}
    - \frac{3317\,\gamma_{\rm E}}{126}
    + \frac{9155185261\,\eta}{548674560}
    - \frac{3977\,\pi^{2}\,\eta}{1152}
    - \frac{5732473\,\eta^{2}}{1306368}
    - \frac{3090307\,\eta^{3}}{139968}
     \\  
    + \frac{87419\,\ln(2)}{1890}
    - \frac{26001\,\ln(3)}{560}
    - \frac{3317}{252}\,\ln\!\Biggl(16 v_0^{2}\Biggr)
    \Biggr) v_0^{6} + \Biggl(\frac{303754441}{56045232}+ \frac{114114347\,\eta}{14011308} 
    + \Biggl(\frac{860536331353}{56493593856}
    - \frac{995744053\,\eta}{147118734}
     \\  
    - \frac{22480526359\,\eta^{2}}{504407088}\Biggr)  v_0^{2} + \Biggl( \frac{114515424257}{4035256704} 
    + \frac{43021108819}{1008814176} \eta  \Biggr) \pi v_0^{3} + \Biggl( -\,\frac{362455290477691}{170836627820544}
    - \frac{1828270052831045\,\eta}{42709156955136}
     \\ 
    + \frac{8331748534489\,\eta^{2}}{508442344704}
    + \frac{2071745969785\,\eta^{3}}{18158655168} \Biggr)v_0^{4} \Biggr)v^2 +
    \Biggl(- \frac{765436643}{64376280}\pi 
    + \Biggl(-\frac{2168482009619}{64891290240} + \frac{150791018671 \eta}{2317546080} \Biggr)\pi v_0^{2} 
     \\  
    -\frac{288569614411}{4635092160} \pi^{2} v_0^{3}\Biggr)v^3 
    + \Biggl(\frac{544449123560815}{36777329600256}
    + \frac{259494480355\,\eta}{5916558816}
    + \frac{545503567565\,\eta^{2}}{15636619728} +      \Biggl(\frac{1542424367047788895}{37071548237058048}
     \\ 
    + \frac{55946348210268175\,\eta}{1323983865609216} 
    - \frac{33395506383635\,\eta^{2}}{235249443072}
    - \frac{107464202810305\,\eta^{3}}{562918310208} \Biggr)v_0^{2} \Biggr)v^4 + \Biggl(
    - \frac{138415580623567}{2100855521520}
     \\  
    - \frac{3736280491255 \eta}{60024443472}
    \Biggr)\pi v^{5}
    + \Biggl(-\frac{10703396867473640999267}{26006825931609600000} 
    + \frac{3808074467687\,\pi^{2}}{46350921600}
    + \frac{1349929441\,\gamma_{\rm E}}{23667750}
     \\ 
    + \frac{4334671685651227\,\eta}{4358077240320}
    - \frac{376528789\,\pi^{2}\,\eta}{12117888}
    + \frac{8865915744977417\,\eta^{2}}{56291831020800}
    + \frac{898457085967259\,\eta^{3}}{8443774653120}
    + \frac{1582619773 \ln(2)}{71003250}
     \\  
    - \frac{461751759 \ln(3)}{21038000} + \frac{1349929441 \ln(16 v^2)}{47335500} \Biggr)v^6 \Biggr\} + \frac{2983 e_{\rm 0}^2 v^{5/3} }{816 v_{\rm 0}^{5/3}}  \Biggl \{1 + \Biggl(\frac{42431}{6384} - \frac{2417 \eta}{684}\Biggr) v_{\rm o}^2 + \frac{2251 \pi v_{\rm 0}^3}{1710} + \Biggl(\frac{943173641}{57915648} 
     \\  
    - \frac{26687 \eta}{3024} + \frac{123719 \eta^2}{24624}\Biggr) v_{\rm 0}^4 + \Biggl(-\frac{4740011}{287280} + \frac{817781  \eta}{215460}\Biggr) \pi v_{\rm 0}^5 + \Biggl(-\frac{51727624818523}{486491443200} + \frac{16107017 \pi^2}{984960} + \frac{23861 \gamma_{\rm E}}{1710} 
     \\ 
    + \frac{10675715266147 \eta}{10424816640} 
    - \frac{969773 \pi^2 \eta}{21888} - \frac{109105939 \eta^2}{2757888} - \frac{3102349 \eta^3}{2659392} + \frac{170237 \ln(2)}{1890} - \frac{962037 \ln(3)}{10640} + \frac{23861 \ln(16 v_{\rm 0}^2)}{3420}\Biggr) v_{\rm 0}^6 
     \\ 
    +  \Biggl(\frac{6955261}{2215584} + \frac{436441 \eta}{79128} + \Biggl(\frac{295118679491}{14144288256} + \frac{4843127297 \eta}{189432432} - \frac{1054877897 \eta^2}{54123552}\Biggr) v_{\rm 0}^2 + \Biggl(\frac{15656292511 \pi}{3788648640} + \frac{982428691 \pi \eta}{135308880}\Biggr) v_{\rm 0}^3 
     \\ 
    + \Biggl(\frac{6560018841475301}{128316983058432} + \frac{284678952641693 \eta}{4582749394944} - \frac{1795086684817 \eta^2}{54556540416} + \frac{53996044079 \eta^3}{1948447872}\Biggr) v_{\rm 0}^4\Biggr)v^2 + \Biggl(-\frac{1114537 \pi}{141300} 
     \\ 
    + \Biggl(-\frac{47290919447}{902059200} + \frac{2693835929 \eta}{96649200}\Biggr) \pi v_{\rm 0}^2 - \frac{2508822787 \pi^2 v_{\rm 0}^3}{241623000}\Biggr) v^3 +  \Biggl(\frac{377620541}{107433216} + \frac{561233971 \eta}{31334688} + \frac{36339727 \eta^2}{2238192} 
     \\ 
    + \Biggl(\frac{16022817175171}{685853650944} + \frac{383924201690765 \eta}{3600731667456} + \frac{1912787853197 \eta^2}{42865853184} - \frac{87833120159 \eta^3}{1530923328}\Biggr) v_{\rm 0}^2\Biggr)v^4 + \Biggl(-\frac{131697334 }{8456805} 
     \\ 
    - \frac{268652717 \eta}{9664920}\Biggr) \pi v^5 +  \Biggl(-\frac{204814565759250649}{1061268280934400} + \frac{365639621 \pi^2}{13022208} + \frac{12483797 \gamma_{\rm E}}{791280} + \frac{34787542048195 \eta}{137827049472} - \frac{8764775 \pi^2 \eta}{1446912} 
     \\ 
    + \frac{80353703837 \eta^2}{1640798208} + \frac{5885194385 \eta^3}{175799808} + \frac{89383841 \ln(2)}{2373840} - \frac{26079003 \ln(3)}{703360} + \frac{12483797 \ln(16 v^2)}{1582560} \Biggr)v^6
    \Biggr\}\Biggr] \Biggr]        
        \tag{A3}
    \label{eq:observer_frame_orbital_phase}
\end{multline}

\section{Full Phase correction}\label{app:full_phcor}

\begin{multline}
\Delta \Psi  =  \frac{25 M \Gamma}{65536 v^{13} \eta ^2}  \Biggl[1 
+ \Biggl(\frac{743}{126} + \frac{22 \eta}{3}  \Biggr) v^2 - \frac{64 \pi v^3}{5}  
+  \Biggl(\frac{1755623}{84672}+\frac{32633 \eta }  {756} + \frac{367 \eta ^2}{12} \Biggr) v^4 +\Biggl(-\frac{20807 }{210}-\frac{574  \eta}{15}  \Biggr) \pi v^5
\\  
+  \Biggl(-\frac{28907482848623}{35206617600}+\frac{9472 \pi ^2}{75} + \frac{13696 \gamma_{\rm E} }{105} +\frac{3311653861 \eta }{1524096} -\frac{451 \pi ^2 \eta }{6} +\frac{2030687 \eta ^2}{18144} +\frac{66287 \eta ^3}{648} 
+ \frac{6848 \ln (16 v^2)}{105} \Biggr) v^6
 \\  
 +\Biggl(-\frac{158992529 }{317520}-\frac{1015907 \eta }{1890} + \frac{419 \eta ^2}{945} \Biggr)  \pi v^7 -\frac{314 e_{\rm 0}^2 v_{\rm 0}^{19/3}}{43 v^{19/3}} \Biggl\{ 1+ \Biggl(\frac{2833}{1008}-\frac{197 \eta }{36}\Biggr) v_{\rm 0}^2 + \frac{377 \pi  v_{\rm 0}^3}{72} + \Biggl( -\frac{1193251}{3048192} -\frac{66317 \eta }{9072}
\\  
 + \frac{18155 \eta ^2}{1296}\Biggr) v_{\rm 0}^4 
+\Biggl(\frac{764881 }{90720}-\frac{949457 \eta }{22680} \Biggr)  \pi v_{\rm 0}^5
+  \Biggl(\frac{26531900578691}{168991764480}+\frac{122833 \pi ^2}{10368}-\frac{3317 \gamma_{\rm E} }{126}
+\frac{9155185261 \eta }{548674560}-\frac{3977 \pi ^2 \eta }{1152}
\\  
-\frac{5732473 \eta ^2}{1306368}-\frac{3090307 \eta ^3}{139968} +\frac{87419 \ln (2)}{1890} -\frac{26001 \ln (3)}{560}
-\frac{3317}{252} \ln \left(16 v_{\rm 0}^2\right) 
\Biggr)v_{\rm 0}^6 + \Biggl(\frac{4979581}{836496} + \frac{1870727 \eta }{209124}
\\ 
+\Biggl(\frac{14107152973}{843187968}-\frac{16323673 \eta }{2195802}-\frac{368533219 \eta ^2}{7528464}\Biggr) v_{\rm 0}^2 +\Biggl(\frac{1877302037}{60227712}+\frac{705264079 \eta }{15056928}
 \Biggr)  \pi  v_{\rm 0}^3 +\Biggl(-\frac{5941890007831}{2549800415232}
\\  
 -\frac{29971640210345 \eta }{637450103808}+\frac{136586041549 \eta ^2}{7588691712}
+\frac{33963048685 \eta ^3}{271024704}
\Biggr) v_{\rm 0}^4 
\Biggr)v^2 +  \Biggl(-\frac{26394367 \pi }{1921680}+ \Biggl(-\frac{74775241711 }{1937053440}
 \\ 
 +\frac{5199690299\eta }{69180480} \Biggr) \pi  v_{\rm 0}^2
  - \frac{9950676359 \pi ^2 v_{\rm 0}^3}{138360960} \Biggr) v^3 +  \Biggl(\frac{9899074973833}{548915367168}+\frac{4718081461 \eta }{88306848} +\frac{9918246683 \eta ^2}{233382384} 
 \\   
 +\Biggl(\frac{28044079400868889}{553306690105344}
 +\frac{1017206331095785 \eta }{19760953218048}-\frac{40681798685519 \eta ^2}{235249443072}-\frac{1953894596551 \eta ^3}{8401765824} 
 \Biggr) v_{\rm 0}^2 \Biggr)v^4 
  \Biggr) 
  \\ 
  + \Biggl(-\frac{10647352355659 }{125424210240}-\frac{287406191635  \eta }{3583548864}\Biggr) \pi  v^5
 + \Biggl( -\frac{1516082854046812099781}{2717131067481600000}+\frac{77715805463 \pi ^2}{691804800}+  \frac{192847063 \gamma_{\rm E} }{2472750} 
 \\  
 +\frac{4334671685651227 \eta }{3187250519040}-\frac{7684261 \pi ^2 \eta }{180864}
 +\frac{1266559392139631 \eta ^2}{5881236076800}
 +\frac{18335858897291 \eta ^3}{126026487360}  
 +\frac{226088539 \ln(2)}{7418250} 
  \\   
 -\frac{65964537 \ln (3)}{2198000} + \frac{192847063 \ln(16 v^2)}{4945500}\Biggr) v^6 \Biggr\}
 + \frac{2983 e_{\rm 0}^2 v^{5/3}}{731 v_{\rm 0}^{5/3}} \Biggl\{1 + \Biggl(\frac{42431}{6384} - \frac{2417 \eta }{684}\Biggr) v_{\rm 0}^2+\frac{2251 \pi  v_{\rm 0}^3}{1710} + \Biggl(\frac{943173641}{57915648}-\frac{26687 \eta }{3024}
   \\ 
   +\frac{123719 \eta ^2}{24624}\Biggr) v_{\rm 0}^4
 +\Biggl(-\frac{4740011}{287280}+\frac{817781  \eta }{215460}\Biggr)  \pi v_{\rm 0}^5  +  \Biggl(-\frac{51727624818523}{486491443200}+\frac{16107017 \pi ^2}{984960}+\frac{23861 \gamma_{\rm E} }{1710}+\frac{10675715266147 \eta }{10424816640}
 \\ 
 -\frac{969773 \pi ^2 \eta }{21888}
 -\frac{109105939 \eta ^2}{2757888}-\frac{3102349 \eta ^3}{2659392}
 +\frac{170237 \ln (2)}{1890}-\frac{962037 \ln (3)}{10640}+\frac{23861 \ln \left(16 v_{\rm 0}^2\right)}{3420}\Biggr)v_{\rm 0}^6 +  \Biggl(\frac{299076223}{81976608} 
  \\ 
  + \frac{18766963 \eta }{2927736}
 +\Biggl(\frac{12690103218113}{523338665472}+\frac{208254473771 \eta }{7008999984}-\frac{45359749571 \eta ^2}{2002571424}\Biggr) v_{\rm 0}^2+\Biggl( \frac{673220577973 }{140179999680} +\frac{42244433713 \eta }{5006428560} 
 \Biggr) \pi v_{\rm 0}^3
  \\  
    + \Biggl(\frac{282080810183437943}{4747728373161984}
  +\frac{12241194963592799 \eta }{169561727612928}
 -\frac{77188727447131 \eta ^2}{2018591995392}+\frac{2321829895397 \eta ^3}{72092571264} \Biggr) v_{\rm 0}^4 \Biggr)v^2 
  \\ 
 +  \Biggl(-\frac{2819123 \pi }{282600} 
 +\Biggl(-\frac{119618208013 }{1804118400}+\frac{6813820291  \eta }{193298400}\Biggr) \pi v_{\rm 0}^2
 -\frac{6345845873 \pi ^2 v_{\rm 0}^3}{483246000}\Biggr)v^3 + \Biggl(\frac{16237683263}{3330429696}+\frac{24133060753 \eta }{971375328}
  \\ 
 +\frac{1562608261 \eta ^2}{69383952}+\Biggl(\frac{688981138532353}{21261463179264}+\frac{16508740672702895 \eta }{111622681691136}+\frac{82249877687471 \eta ^2}{1328841448704}-\frac{3776824166837 \eta ^3}{47458623168} \Biggr) v_{\rm 0}^2\Biggr) v^4 
  \\ 
 + \Biggl(-\frac{2831492681 }{118395270}-\frac{11552066831   \eta }{270617760}\Biggr) \pi v^5 +  \Biggl(-\frac{43603153867072577087}{132658535116800000}+\frac{15722503703 \pi ^2}{325555200}+\frac{536803271 \gamma_{\rm E} }{19782000}
 \\ 
 +\frac{299172861614477 \eta }{689135247360}-\frac{15075413 \pi ^2 \eta }{1446912}+\frac{3455209264991 \eta ^2}{41019955200}+\frac{50612671711 \eta ^3}{878999040}+ \frac{3843505163 \ln(2)}{59346000}
 \\ 
 -\frac{1121397129 \ln(3)}{17584000}  + \frac{536803271 \ln(16 v^2)}{39564000}\Biggr) v^6
 \Biggr\} \Biggr]
 \tag{A4}
    \label{eq:full_ph_cor}
\end{multline}

\section{Resolving the discrepancy between this work and~\cite{Roy:2026mco}}
\label{sec: rep_our_from_soumen}
To derive the correct eccentric phase correction expression from that derived in~\cite{Roy:2026mco}, we follow the procedure described in Section (III) of~\cite{Ansari:HM_2026}. The only difference being that here we use eccentric expressions. Using Newtonian-order eccentric time and orbital phase expressions given by Eqs. (6.7b) and (6.6b) of ~\cite{Moore:2016qxz} respectively, 
\[ t_{\rm src} = t_{\rm c} -\frac{5 M }{256 \eta v_{\rm src}^8} \Biggl(1 - \frac{157 e_{0}^2}{43} \Biggl(\frac{v_{\rm 0src}}{v_{\rm src}}\Biggr)^{19/3} \Biggr)\] 
and 
\[\phi = \phi_{\rm c} -\frac{1}{32 \eta v_{\rm src}^5} \Biggl(1-\frac{785 e_{0}^2}{272} \Biggl(\frac{v_{\rm 0src}}{v_{\rm src}}\Biggr)^{19/3} \Biggr)\]
along with $f = \frac{v^3}{\pi M}$ in Equation (25) of ~\cite{Ansari:HM_2026}, we get
\begin{multline}
    \Psi_{\rm tot} = \frac{2v^3}{M} \Biggl(t_{\rm c} - \frac{5 M }{256 \eta v_{\rm src}^8} \Biggl(1 - \frac{157 e_{0}^2}{43} \Biggl(\frac{v_{\rm 0src}}{v_{\rm src}}\Biggr)^{19/3} \Biggr)\Biggr) - 2 \Biggl(\phi_{\rm c} - \frac{1}{32 \eta v_{\rm src}^5} \Biggl(1-\frac{785 e_{0}^2}{272} \Biggl(\frac{v_{\rm 0src}}{v_{\rm src}}\Biggr)^{19/3} \Biggr) \Biggr) \\ -\frac{\pi}{4} + \frac{\Gamma v^3}{M} \Biggl(-\frac{5 M }{256 \eta v_{\rm src}^8} \Biggl(1 - \frac{157 e_{0}^2}{43} \Biggl(\frac{v_{\rm 0src}}{v_{\rm src}}\Biggr)^{19/3} \Biggr)\Biggr)^2
    \label{eq: D1}
\end{multline}
Substituting $v_{\rm src} = v \left(1 + \Gamma (t_{\rm u} - t_{\rm c}) \right)^{1/3}$, \ \ $v_{\rm 0src} = v_{0} \left(1 + \Gamma (t_{\rm 0u} - t_{\rm c}) \right)^{1/3} $ and, to the leading order in $\Gamma$, $\Gamma t_{\rm u} = \Gamma t_{\rm o}$, in Equation ~\eqref{eq: D1} results in
\begin{equation}
    \Psi_{\rm tot} = 2 \pi f t_{\rm c} - 2 \phi_{\rm c} -\frac{\pi}{4} + \frac{3}{128 \eta v^5} \Biggl(1 - \frac{2355 e_{0}^2}{1462} \Biggl(\frac{v_0}{v}\Biggr)^{19/3} \Biggr) + \frac{25 M \Gamma}{65536 \eta^2 v^{13}} \Biggl(1 - \frac{314 e_{0}^2}{43} \Biggl(\frac{v_0}{v}\Biggr)^{19/3} + \frac{2983 e_{0}^2}{731} \Biggl(\frac{v}{v_0}\Biggr)^{5/3} \Biggr).
    \label{eq: D2}
\end{equation}
Here, the term without $\Gamma$ is the Newtonian-order eccentric Fourier phase, given in Equation (6.26) of ~\cite{Moore:2016qxz}, and the $\Gamma$ term is the phase correction, which is consistent with Equation~\eqref{eq: phase_cor}.

\section{Additional Figures}\label{app: add_fig}
\begin{figure*}[h]
    \centering
    \includegraphics[width=0.475\linewidth]{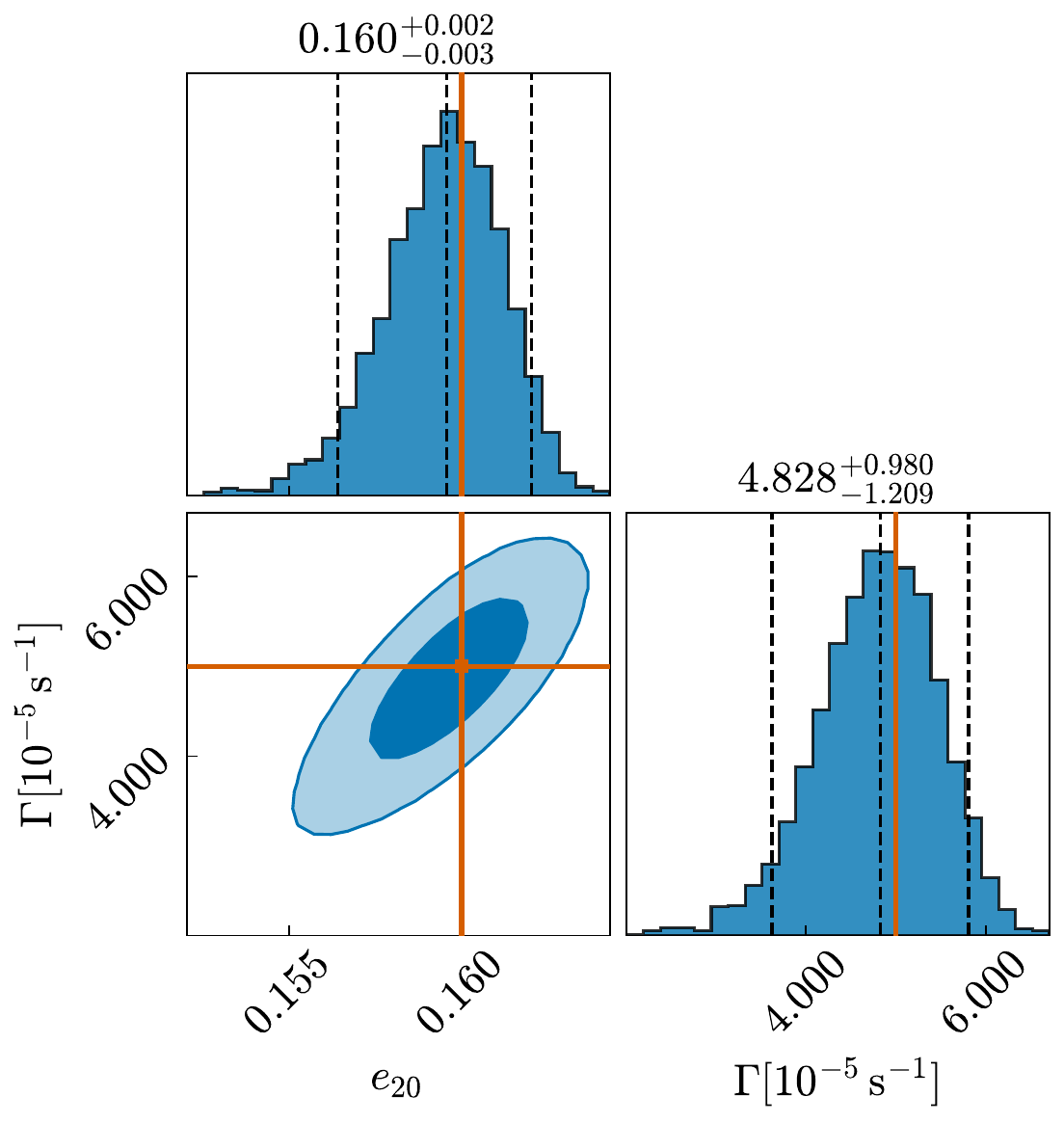}
    \includegraphics[width=0.475\linewidth]{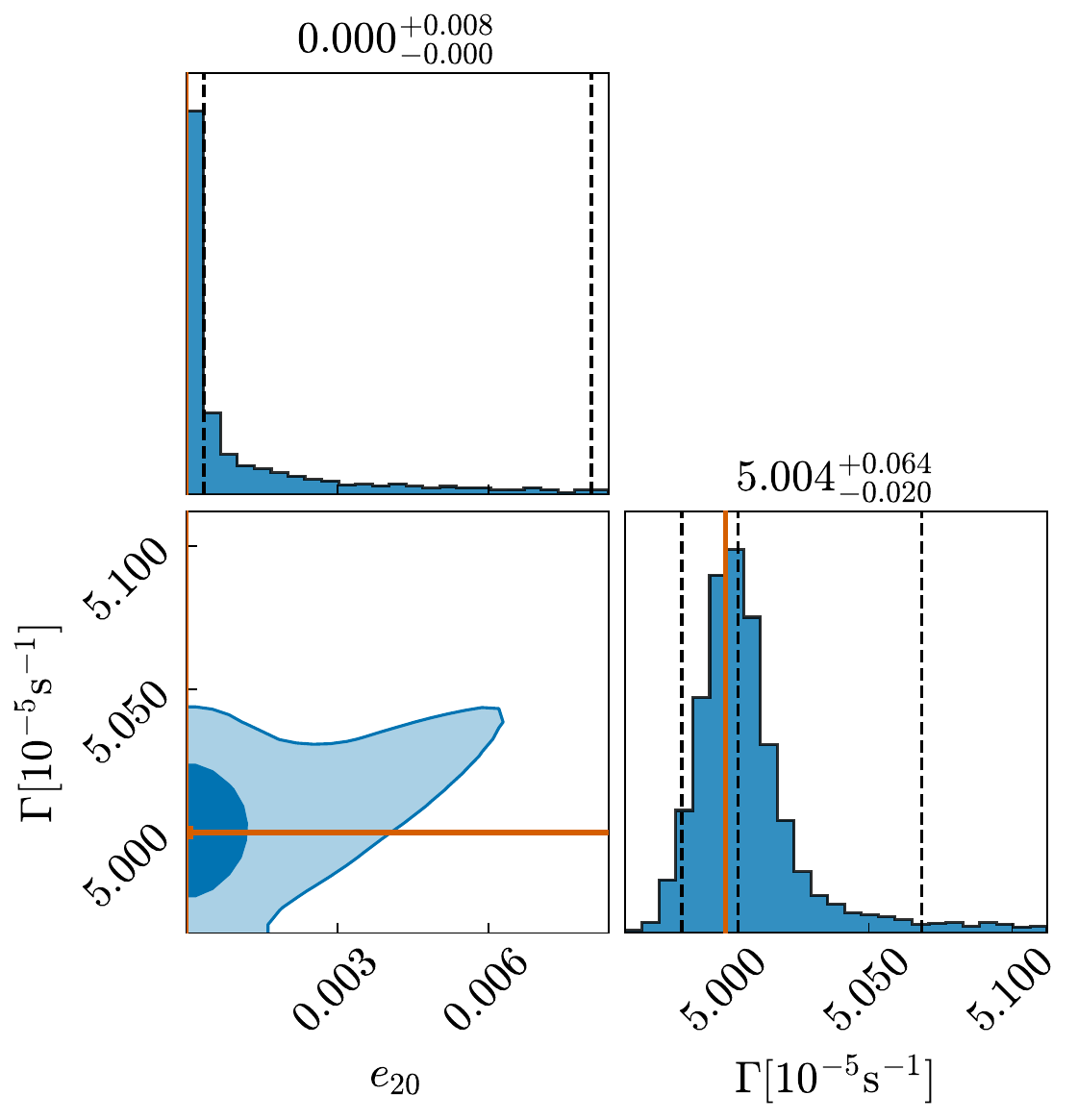}
    \label{fig: self_consist_positive_losa}
    \caption{{\bf Removing the Bias:} The {\it left panel} shows the joint posteriors of $e_{20}$ and $\Gamma$ for an accelerating eccentric CBC, with $\{e_{20} = 0.16,\, \Gamma = 5 \times 10^{-5} \, \rm s^{-1} \}$, when the injection and recovery waveforms are the same, while the {\it right panel} shows the same for an accelerating quasi-circular CBC with $\{\Gamma = 5 \times 10^{-5} \, \rm s^{-1} \}$.}
\end{figure*}

Figure~\ref{fig: self_consist_positive_losa} shows the joint posteriors of $e_{20}$ and $\Gamma$ for an accelerating eccentric CBC ({\it left panel}) and an accelerating quasi-circular CBC ({\it right panel}) for a positive LOSA. Notice that for both injections, parameters are consistent with their injected values and lie within the corresponding $90\%$ CIs.

\end{document}